\documentclass [12pt,fleqn]{article}

\usepackage{verbatim}
\usepackage{graphicx}
\usepackage{amssymb}
\usepackage{amsmath}
\usepackage{lscape}
\usepackage{color}

\newtheorem {lemma} {Lemma}

\newtheorem {proposition} {Proposition}
\newtheorem{applemma}{Lemma}[section]

\newenvironment{apabib}{
  \noindent
{\large\bf References} \vspace*{-6mm} %orig: \vspace*{-3mm}
\begin{list}{}{
  \topsep          0mm
  \leftmargin     10mm
  \parsep         2mm %orig: 2.3mm
  \listparindent -10mm}
  \item[]\strut\par}{
\end{list}}

\newcommand {\topcaption}{%
             \setlength{\abovecaptionskip}{0pt}%
             \setlength{\belowcaptionskip}{10pt}%
             \caption}

\newcounter{excount}

\newcounter{asscount}
\newenvironment{assumption}
{\refstepcounter{asscount}\bigskip\noindent\textsc{Assumption~\arabic{asscount}}}
{\smallskip}

\newcounter{defcount}

\newcommand{\prf}{\noindent {\bf Proof}:\ \ } %%for beginning proofs
\newcommand{\eprf}{$\;\rule{1.5mm}{3mm}$ \smallskip}

\newcommand {\reals}{\mathbb{R}}
\newcommand {\rar}{\rightarrow}

\renewcommand{\mid}{|\,}
\newcommand{\TP}{TP}
\newcommand{\FP}{FP}

\renewcommand{\P}{\mathbb{P}}

\begin{document}
%\setcounter{page}{1}
%\vspace*{15mm}
%\begin{center}

%{\Large \bf  Inference for ROC Curves Based on Estimated Predictive
%Indices}

%\vspace*{20mm} {\large\bf Yu-Chin Hsu$^{\ast}$}

%\vspace*{3mm}
%Institute of Economics, Academia Sinica;\\
%Department of Finance, National Central University; \\
%Department of Economics, National Chengchi University;\\
%CRETA, National Taiwan University

%\vspace*{10mm} {\large\bf Robert P.\ Lieli$^{\dagger}$}
%
%\vspace*{3mm}
%Department of Economics, Central European University, Vienna\\

%\vspace*{20mm} {\bf (Preliminary: please do not cite or quote
%without permission.)}\\
%First version: March 26, 2014 \\
%April, 2021
%
%\vspace*{3mm}

%\end{center}

%\vfill \noindent $^{\ast}${\footnotesize Email: ychsu@econ.sinica.edu.tw } \\
%$^{\dagger}${\footnotesize  Email: lielir@ceu.edu } \\

\title{Inference for ROC Curves Based on Estimated Predictive Indices\thanks{
%We are grateful to *** for valuable comments. 
Yu-Chin Hsu gratefully acknowledges the research support from Ministry of Science and Technology of Taiwan (MOST 110-2628-H-001-007), Academia Sinica Investigator Award (AS-IA-110-H01) and Center for Research in Econometric Theory and Applications (110L9002) from the Featured Areas Research Center Program within the framework of the Higher Education Sprout Project by the Ministry of Education of Taiwan.}}

\author{Yu-Chin Hsu\thanks{Institute of Economics, Academia Sinica; Department of
Finance, National Central University; Department of Economics, National Chengchi University; CRETA, National Taiwan University. E-mail: ychsu@econ.sinica.edu.tw.}
\and Robert P. Lieli\thanks{Department of Economics and Business, Central European University, Vienna.  E-mail: lielir@ceu.edu.} }

\maketitle

\begin{abstract}

We provide a comprehensive theory of conducting in-sample statistical inference about receiver operating characteristic (ROC) curves that are based on predicted values from a first stage model with estimated parameters (such as a logit regression). The term ``in-sample'' refers to the practice of using the same data for model estimation (training) and subsequent evaluation, i.e., the construction of the ROC curve. We show that in this case the first stage estimation error has a generally non-negligible impact on the asymptotic distribution of the ROC curve and develop the appropriate pointwise and functional limit theory. We propose methods for simulating the distribution of the limit process and show how to use the results in practice in comparing ROC curves.
%construct uniform confidence bands, compare areas under two (arbitrarily correlated) ROC curves, or test for (partial) dominance.\medskip 

\noindent \textit{Keywords:} classification, ROC curve, uniform asymptotics, estimation effect 

\noindent \textit{JEL codes: C25, C38, C46, C52} 
\end{abstract}

\newpage

\section{Introduction}

Binary prediction or classification is a fundamental problem in statistics and machine learning with applications in many scientific disciplines including economics. The predictive ability of statistical models used for binary classification is frequently evaluated through the receiver operating characteristic (ROC) curve, designed to summarize the tradeoffs between the probability of a true positive prediction (vertical axis) and a false positive prediction (horizontal axis) as one combines a predictive index with a varying classification threshold.
%Another way to think about this curve is that it measures the extent to which the density of the predictive index conditional on a negative outcome ($f_0$) overlaps with the corresponding density conditional on a positive outcome ($f_1$).  
Though its origins are in the signal detection and medical diagnostics literature, in recent years ROC analysis has become increasingly common in financial and economic applications as well (e.g., Anjali and Bossaerts 2014; Bazzi et al.\ 2021; Bonfim et al.\ 2021; Berge and Jorda 2011; 
%Carole and Putnins 2014; 
%Jorda and Taylor 2011; 
Kleinberg et al.\ 2018; Lahiri and Wang 2013; Lahiri and Yang 2018; McCracken et al.\ 2021; Schularik and Taylor 2012 and many others).

While there is a large literature on the statistical properties of empirical ROC curves, the standard distributional theory assumes that the signal or predicitive index used for classification is either directly observed---it is ``raw data''---or that it is a fixed function of raw data. However, if the signal itself is generated from an underlying regression model with estimated coefficients, conducting in-sample inference about ROC curves based on the traditional theory can be highly misleading. For instance, Demler et al.\ (2012) point out that the standard DeLong et al.\ (1988) test for comparing AUCs for different (but potentially correlated) signals can lead to flawed inference if the signals come from nested models with estimated coefficients.\footnote{AUC stands for ``area under [the ROC] curve''. It is an overall performance measure for binary prediction models. AUC=1/2 corresponds to no predictive power.} Similarly, Lieli and Hsu (2019) demonstrate that the asymptotic normality results in Bamber (1975) are inappropriate for testing AUC=1/2 for models with estimated parameters. 

The central contribution of this paper is the development of a general functional limit theory for the empirical ROC curve that takes the pre-estimation effect into account. Regarding the ROC curve as a random function defined over the [0,1] interval, we provide a uniform influence function representation theorem, and show that the difference between the empirical and population ROC curves converges weakly to a mean zero Gaussian process with a given covariance structure at the parametric rate. These results constitute a non-trivial extension of the functional limit result in Hsieh and Turnbull (1996), who work under the assumption that the observations available on the predictive index are i.i.d.\ conditional on the outcome. (If the predictive index depends on coefficients estimated in-sample, this assumption is no longer valid, as the parameter estimates depend on all data points.) Our results not only allow for the construction of a uniform confidence band for the ROC curve but also facilitate model selection through handling virtually any comparison between two correlated ROC curves (e.g., testing dominance or partial dominance; testing the difference between AUCs or partial AUCs, etc.) In terms implementation, we propose two methods to simulate the limiting distribution of the empirical ROC curve, one of which is the weighted bootstrap by Ma and Kosorok (2005). Although some type of bootstrap procedure would be a natural way to approach the pre-estimation problem in practice even without a theory, our results provide rigorous justification and guidance for doing this. 

%it is a natural inclination for practitioners may well resort to some type of bootstrap procedure to handle pre-estimation problems our theory provides    

A second contribution of the theory developed in this paper is that it provides insight into what determines the impact of the first stage estimation error on the asymptotic distribution of the ROC curve. The derivatives of the true and false positive prediction rates with respect to the first stage model parameters play a central role --- if these gradients vanish at the pseudo-true parameter values, then so does the estimation effect. Nevertheless, the gradients also depend on the classification threshold and will not generally be negligible along the entire ROC curve. For associated functionals, it is the gradient of the functional that drives the estimation effect. Some functionals, e.g., the area under the curve, have the property that they are maximal when the predictive index is given by $p(X)=P(Y=1|X)$, the conditional probability of a positive outcome $Y$ given the covariates $X$. For such functionals the estimation effect is negligible when the first stage model is correctly specified for $p(X)$
%, this function depends non-trivially on $X$, 
and the first stage estimator converges at the root-n rate. Nevertheless, the first stage estimation error will generally affect the asymptotic distribution under misspecification; thus, our results facilitate robust inference.

There are additional technical contributions that are more subtle. In employing standard empirical process techniques to derive our results, we make most of the fact that the population and sample ROC curves are invariant to monotone increasing transformations of the predictive index. This observation allows us to leverage powerful assumptions that may seem restrictive at first glance. In particular, we use the assumption that the density $f_0$ of the predictive index conditional on $Y=0$ is bounded away from zero to derive various uniform approximations. The problem is that even if the individual predictors have densities bounded away from zero (already a big \textit{if}), the predictive index may not share this property, as it often involves a linear combination of the predictors.\footnote{Think of the sum of two independent uniform[0,1] random variables or the central limit theorem for that matter.} Nevertheless, one can always find a strictly increasing transformation of the predictive index, say, the probability integral transform, so that the post-transform $f_0$ will be greater than some $\epsilon>0$ across the whole support. What matters for the asymptotic theory is the properties of the likelihood ratio $f_1/f_0$, which is invariant to monotone increasing transformations (here $f_1$ is the conditional density of the predictive index given $Y=1$). In particular, uniform inference is possible only for parts of the ROC curve that are generated by thresholds falling into some interval $[c_L,c_U]$ over which $f_1/f_0$ is bounded and bounded away from zero.\footnote{That such an interval exists is a weak assumption; that it coincides with the support of $f_0$ is a much stronger one.} But given the properties of the likelihood ratio, one is free to assume the theoretically most convenient scenario about the individual density $f_0$ that is achievable through monotone transformations, even if this transformation is not implemented or even identified.

%With proper in-sample theory in hand, there remains little reason to conduct out-of-sample inference on ROC curves. The external validity of inferences about predictive ability is not enhanced by merely holding out part of the same sample for evaluation. If the concept of overfit is extended to searches over different model specifications, both in-sample tests and out-of-sample tests are affected---a point forcefully made by Inoue and Kilian (2004). Hence, the choice comes down to power, where in-sample tests are expected to be better.

We must also point out some technical limitations of the paper. First, we do not allow for serial dependence in the data, precluding time series applications such as the evaluation of recession forecasting models. Nevertheless, our proofs rely mostly on high level conditions; specifically, the asymptotically linear representation of the first stage estimator, the stochastic equicontinuity of the empirical process defined by the (pseudo-true) predictive index, and the uniform continuity of some derivatives. Given the availability of these conditions for stationary, weakly dependent time series, we conjecture that our representation results should generalize to this setting with relatively straightforward modifications. However, simulating the asymptotic distribution of the limiting process would require more complex procedures and we do not pursue this extension here. Second, the predictive model evaluated at the pseudo-true parameter values must have strictly positive variance. This is not an innocent assumption in that it rules out a completely uninformative predictive model. For example, our results are not suitable for testing the hypothesis that AUC=1/2; see Lieli and Hsu (2019) for some specialized results in this very non-standard case. More generally, in using our results to compare two ROC curves, the difference of the two influence functions evaluated at the pseudo-true parameter values must have strictly positive variance as well. This condition can be violated when the first stage models are nested, and we are currently working on some test procedures that are applicable in this scenario as well. Finally, we only consider parametric estimators of $p(X)$ as the first stage model; nonparametric estimators that converge slower than the root-n rate are ruled out. 

One might discount the practical relevance of our theoretical results discussed above based on the fact that the first stage estimation problem can be avoided by conducting out-of-sample evaluation. If the ROC curve is constructed over a test sample that is independent of the training sample used to estimate the first stage model, then the asymptotic validity of the standard inference procedures is restored. We acknowledge this point but offer two responses. First, out-of-sample evaluation is costly: it leads to power loss in model comparisons and the potential dependence of the results on the particular split(s) used. In fact, one could argue that out-of-sample evaluation is a necessity forced on practitioners by the fact that it is often very difficult to characterize analytically or in a practically useful way how goodness-of-fit measures behave over the training sample so that one can compensate for overfitting. In this case we do provide such a result. 
%\textcolor{red}{and, as an additional contribution, we also present data-driven Monte Carlo simulations based on Demler et al.\ (2012) to demonstrate the power gains afforded by using in-sample tests of classification performance.} 
Second, apart from dealing with the pre-estimation problem, our results provide a unified framework for conducting uniform inference, and comparing ROC curves estimated over the same sample in virtually any way.  

This work has ties to several strands of the statistics and econometrics literature. We have already cited a number of classic works on the statistical properties of the empirical ROC curve that maintain the assumption of a directly observed signal (Bamber 1975, DeLong et al.\ 1988, Hsieh and Turnbull 1996). It is the last of these papers that is closest to ours; however, the pre-estimation effect is obviously missing from their framework and they actually do not exploit their functional limit result for inference apart from re-deriving the asymptotic normality result of Bamber (1975) for the empirical AUC. Instead, they focus on estimating the ROC curve under an additional ``binormal'' assumption, i.e., when the signal has a normal distribution conditional on both outcomes.

Pre-estimation problems have a long history in the literature; for example, Pagan (1984) studied the distributional consequences of including ``generated regressors'' into a regression model. As mentioned above, Demler et al.\ (2012) pointed out the relevance of the pre-estimation effect in the context of ROC analysis. More generally, our work is related to papers dealing with two-step estimators where the first step involves estimating some nuisance parameter whose sampling variation potentially affects the otherwise well-understood second stage. Abadie and Imbens (2016) is a relatively recent example in the context of matching estimators. 
%\textcolor{blue}{[A-Chin, do you want to add more references on two stage estimation here?]} \textcolor{blue}{[Robert, I am thinking that the literature on pre-estimation is quite standard and large.  Maybe we can just ignore this part?]}

The application of our results in testing for dominance relations and AUC differences across ROC curves has similarities to stochastic dominance tests; see, e.g., Barrett and Donald (2003), Linton, Maasoumi and Whang (2005), Linton, Song and Whang (2010) and Donald and Hsu (2016). Some papers, such as Linton, Maasoumi and Whang (2005) and Linton, Song and Whang (2010), even allow for generated variables in this context.  Finally, the paper speaks indirectly to the forecasting literature on the relative merits of in-sample vs.\ out-of-sample model evaluation (e.g., Inoue and Kilian 2004, Clark and McCracken 2012). The connection lies in the fact that we extend the scope of in-sample evaluation methods in binary prediction. 
%\textcolor{red}{and also show that there are gains from doing so.}  

The rest of the paper is organized as follows. Section \ref{sec: binpred} sets up the prediction framework and introduces the ROC curve along with some of its basic properties. Section \ref{sec: pointwise results} discusses the estimation effect in detail and presents pointwise (fixed-cutoff) asymptotic results. The functional limit theory is contained in Section \ref{sec: uniform results}. In Section \ref{sec: inference} we show how to use the abstract results for conducting inference about the ROC curve; we discuss dominance testing, AUC comparisons, etc. Section \ref{sec: simulations} presents Monte Carlo results highlighting the impact of first stage estimation on the distribution of the ROC curve.  
%and the power gains from in-sample rather than out-of-sample evaluation. 
Section~\ref{sec: concl} concludes.

\section{Making and evaluating binary predictions}\label{sec: binpred}

\subsection{Cutoff rules and the ROC curve}

Let $Y\in\{0,1\}$ be a Bernoulli random variable representing some outcome of interest and $X$ be a $k\times 1$ vector of covariates (predictors). We consider point forecasts (classifications) of $Y$ that are constructed by combining a scalar predictive index $G(X)$ with a suitable cutoff (threshold) $c$. More specifically, the prediction rule for $Y$ is given by
\begin{equation}\label{eq: cutoff rule}
    \hat Y(c)=1[G(X)>c],
\end{equation}
where $1[.]$ is the indicator function. The role of the function $G: \reals^k\rar\reals$ is to aggregate the information that the predictors contain about $Y$ while the choice of $c$ governs the use of this information. With $G$ and $c$ unrestricted, (\ref{eq: cutoff rule}) represents a very general class of prediction rules. 

In many binary prediction problems there is also a loss function $\ell(\hat y, y)$ that describes the cost of predicting $Y=\hat y$ when the realized outcome is $Y=y$. If the decision maker's objective is to minimize expected loss conditional on $X$, the optimal choice of $G$ and $c$ is determined as follows. Given the observed value of $X$, the optimal point forecast of $Y$ solves \begin{equation}\label{exp loss min}
    \min_{\hat y\in\{0,1\} } E[\ell(\hat y, Y)|X]=\min_{\hat y\in\{0,1\} }\big[\ell(\hat y, 1)p(X)+\ell(\hat y, 0)(1-p(X))\big],
\end{equation}
where $p(X)=\P(Y=1|X)$ is the conditional probability of $Y=1$ given $X$. Adopting the normalization $\ell(0,0)=\ell(1,1)=0$ and assuming $\ell(0,1)>0$ and $\ell(1,0)>0$, it is straightforward to verify that the optimal prediction rule is given by
\begin{equation}\label{eq: opt dec rule}
    \hat Y^*(c_\ell)=1[p(X)>c_\ell],
\end{equation}
where $c_\ell=\ell(1,0)/[\ell(0,1)+\ell(1,0)]$.\footnote{In case $p(X)=c_\ell$, which is often a zero probability event, the decision maker is indifferent between predicting 0 or 1. The formula stated above arbitrarily specifies $\hat Y=0$ in this case.} 

Equation (\ref{eq: opt dec rule}) reveals that the optimal predictive index is $p(X)$ regardless of $\ell$ while the optimal choice of $c$ is fully determined by $\ell$ (specifically, by the relative cost of a false alarm versus a miss). This simple observation motivates a two-step empirical strategy in binary prediction.\footnote{See Elliott and Lieli (2013) for an alternative approach where the decision rule (\ref{eq: opt dec rule}) is estimated in a single step based on a specific loss function.}
 First, one models and estimates $p(X)$ using data on $(Y,X)$; a common approach is to specify a parametric model $G(X,\beta)$ for $p(X)$ and to estimate it by maximum likelihood (logistic regression is a leading example). In the second step a point forecast is obtained by combining the estimated conditional probability $G(X,\hat\beta)$ with a suitable cutoff, which depends on the forecaster's or forecaster user's preferences. Thus, there is a separation between the construction of the predictive index, representing the objective information available to the forecaster, and the use of that information, governed by the loss function.
%\section{The population ROC curve}

%The two step approach to binary prediction described above separates the construction of the predictive index from the forecaster's (or forecast user's) preferences---it is the same information, the estimated probability $G(X,\hat\beta)$, that is reported to any forecaster. The receiver operating characteristic (ROC) curve evaluates the quality of this information simultaneously for all possible loss functions, without committing to any particular one.          

We will now define the population \emph{receiver operating characteristic} (ROC) curve. Let $G(X,\beta)$ be a predictive index with a fixed value of the parameter $\beta$.\footnote{The following definitions do not depend on the parametric structure and generalize immediately to any predictive index $G(X)$. We work with a parametric specification in anticipation of studying the pre-estimation step.} Combined with a cutoff $c$, the resulting prediction rule produces true positive predictions and false positive predictions (false alarms) with the following probabilities:
\begin{eqnarray*}
\TP(c,\beta)&=&\P[\hat Y(c)=1\mid Y=1]=\P[G(X,\beta)>c\mid Y=1]\\
\FP(c,\beta)&=&\P[\hat Y(c)=1\mid Y=0]=\P[G(X,\beta)>c\mid Y=0],
\end{eqnarray*}
where TP and FP  stand for the rate of ``true positive'' and ``false positive'' predictions, respectively. As the cutoff $c$ varies, both quantities change in the same direction; in general, TP can be increased only at the cost of increasing FP as well. The ROC curve traces out all attainable (\FP, \TP) pairs in the $[0,1]\times[0,1]$ unit square, i.e., it is the locus  
\[
\Big\{\big(\FP(c,\beta), \TP(c,\beta)\big): c\in\reals\Big\}.
\]

Intuitively, the ROC curve is a way of summarizing the information content of $G(X,\beta)$ about the outcome $Y$ without committing to any particular cutoff, i.e., loss function. The use of such a forecast evaluation tool is particularly appropriate in situations in which there many potential forecast users with diverse loss functions; see Lieli and Nieto-Barthaburu (2010). It is also clear from the definition that the ROC curve is invariant to strictly monotone transformations of $G(X,\beta)$. 

The ROC curve based on the true conditional probability function $p(X)$ possesses some optimality properties. To state these in a parametric modeling framework, we introduce the following correct specification assumption. 

\begin{assumption}\label{assn: correct spec}
There exists some point $\beta^\circ$ in the parameter space $\mathcal{B}\subseteq\reals^p$ such that $G(X,\beta^\circ)=p(X)$ almost surely. 
\end{assumption}

We state the following result. 

\begin{proposition}\label{prop: p(X) ROC opt}
(i) Given Assumption~\ref{assn: correct spec}, $\beta^\circ$ solves the following maximization problem for any value of $c$:
\begin{equation*}%\label{ROC uc max}
    \max_\beta \big[ (1-c)\pi \TP(c,\beta)-c(1-\pi)\FP(c,\beta)\big],
\end{equation*}
where $\pi=\P(Y=1)$.

(ii) Given Assumption~\ref{assn: correct spec}, define $F^\circ_c= \FP(c,\beta^\circ)$. Then $\beta^\circ$ also solves the following constrained maximization problem for any value of $c$:
\begin{equation*}%\label{ROC uc max}
    \max_\beta \TP(c,\beta)\text{ s.t. }\FP(c,\beta)=F^\circ_c
\end{equation*}
\end{proposition}

\paragraph{Remarks:}

\begin{enumerate}

    %\item Under additional smoothness conditions, part (i) means that if one draws a tangent to the ROC curve of $p(X)$ through the point corresponding to a given value of $c$, then this line will have slope $c(1-\pi)/(1-c)\pi$. Furthermore, for any value of $c$, the tangent line to any 

    \item Part (i) is a consequence of the predictor (\ref{eq: opt dec rule}) solving (\ref{exp loss min}) for any given value of $X$. To see this, note that by the law of iterated expectations and the monotonicity of the expectation operator, $\hat Y^*(c_\ell)$ also solves the unconditional expected loss minimization problem $\min_{\hat Y}E_{XY}[\ell(\hat Y,Y)]$, where the minimization is over all random variables $\hat Y$ that are (measurable) transformations of $X$. It is easy to verify that $E[\ell(\hat Y,Y)]$ can be written as
    \[
    [\ell(0,1)+\ell(1,0)]\cdot\big[ c_\ell(1-\pi)\P(\hat Y=1\mid Y=0)-(1-c_\ell)\pi \P(\hat Y=1\mid Y=1)\big]+\pi\ell(0,1),
    \]
    which immediately implies the result.
    
    \item Part (ii) is a consequence of part (i) and it means that for any given FP rate, it is the ROC curve based on $p(X)$ that achieves the largest possible TP rate. In other words, the ROC curve associated with $p(X)$ weakly dominates any other ROC curve that is constructed based on some index $G(X)$.\footnote{To see this, fix a false positive rate $F_0\in[0,1]$, and find the cutoff $c_0$ that produces $\FP(\beta^\circ, c_0)=F_0$ (for simplicity, assume that exact equality can be achieved). Let $\beta'$ be any other parameter value satisfying $\FP(\beta', c_0)=F_0$. Proposition~\ref{prop: p(X) ROC opt}(i) implies $$(1-c_0)\pi \TP(c_0,\beta^\circ)-c_0(1-\pi)\FP(c_0,\beta^\circ)\ge (1-c_0)\pi \TP(c_0,\beta')-c_0(1-\pi)\FP(c_0,\beta')$$ so that $\TP(c_0,\beta^\circ)\ge \TP(c_0,\beta')$.} 
    
    \item These results are not new; they have appeared in the ROC literature in alternative formulations. See, e.g., Egan (1975) and Pepe (2003, Section 4). 
    
    %and are in some ways analogous to the famous Neyman-Pearson lemma.
    
\end{enumerate}

\subsection{The sample ROC curve and conventional inference}\label{subsec: sample ROC}

Throughout the paper we maintain the assumption that the available data consists of a random sample. More formally: 

\begin{assumption}\label{assn: iid}
The sample $\{(Y_i,X_i)\}_{i=1}^n$ consists of independent and identically distributed observations on the random vector $(Y,X)\in\{0,1\}\times\reals^k$.
\end{assumption}

Given the sample and a \emph{fixed} value of $\beta$, the empirical ROC curve is defined as the locus $\big\{(\widehat{\FP}(c,\beta), \widehat{\TP}(c,\beta)): c\in\reals\big\}\subset [0,1]\times[0,1]$, where 
\begin{eqnarray*}
\widehat{\TP}(c,\beta)&=&\frac{1}{n_1}\sum_{i=1}^n 1[G(X_i,\beta)>c, Y_i=1]\\
\widehat{\FP}(c,\beta)&=&\frac{1}{n_0}\sum_{i=1}^n 1[G(X_i,\beta)>c, Y_i=0],
\end{eqnarray*}
and $n_1=\sum_{i=1}^n Y_i$, $n_0=n-n_1$.

The simplest type of inference about an ROC curve involves constructing (joint) confidence intervals for $\TP(c,\beta)$ and $\FP(c,\beta)$ for one threshold $c$ at a time and for fixed values of the coefficient vector $\beta$. In this case one can use the CLT to arrive at the normal approximations: 
\begin{eqnarray}
&&\sqrt{n}[\widehat{\TP}(c,\beta)-\TP(c,\beta)]\rar_d N\big[0, \TP(1-\TP)/\pi\big] \label{eq: \TP asy dist}\\
&&\sqrt{n}[\widehat{\FP}(c,\beta)-\FP(c,\beta)]\rar_d N\big[0, \FP(1-\FP)/(1-\pi)\big], \label{eq: \FP asy dist}
\end{eqnarray}
where TP is a shorthand for $\TP(c,\beta)$ (and similarly for FP). These results immediately provide asymptotic confidence intervals for TP and FP, and a joint confidence rectangle is also easy to construct due to the independence of $\widehat{\TP}(c,\beta)$ and $\widehat{\FP}(c,\beta)$; see Pepe (2003), Section 2.2.2 for details.\footnote{The two statistics are independent because they are computed from two disjoint sets of observations; namely, the $Y=1$ and $Y=0$ subsamples.} 

%\textcolor{red}{Review Hsieh and Turnbull (1996), Bamber (1975), DeLong (1988)}. 

Furthermore, for fixed $\beta$ one can apply the asymptotic normality results in Bamber (1975) to conduct inference about the AUC, and the DeLong et.\ al.\ (1988) test for comparing the areas under ROC curves based on different (but non-random) values of $\beta$. The nonparametric functional limit result in Hsieh and Turnbull (1996) also applies. %state a general functional limit result (but without a pre-estimation effect) and further study estimation of the ROC curve under the ``binormal'' assumption, i.e., when the signal has a normal distribution conditional on both outcomes. 

%Bamber (1975) uses results on the Mann-Whitney U-statistic to argue that the emprical AUC is asymptotically normal and gives a general formula for the asymptotic variance. Also using U-statistic theory, DeLong et.\ al.\ (1988) propose a test for comparing the areas under ROC curves based on different (but potentially correlated) signals. Hsieh and Turnbull (1996) state a general functional limit result (but without a pre-estimation effect) and further study estimation of the ROC curve under the ``binormal'' assumption, i.e., when the signal has a normal distribution conditional on both outcomes.    

%More generally, inference about various other functionals associated with an ROC curve can be, at least in theory, based on the results by Hsieh and Turnbull (1996). Specifically, they state the Gaussian process that characterizes the asymptotic distribution of the sample ROC curve when viewed as a random function over [0,1], traced out as one varies the cutoff $c$.\footnote{Thus, functionals that depend on a continuum of cutoff values can be considered; see Pepe (2003), Section 4.3 for commonly used examples.} However, these results still only translate to the present setting when the value of the parameter vector $\beta$ is fixed. [CHECK DIFFERENCES MORE CAREFULLY]

\section{In-sample inference: pointwise asymptotics}\label{sec: pointwise results}

We will now develop a comprehensive theory of in-sample inference about individual points on the ROC curve
%, corresponding to fixed values of the cutoff $c$, while 
taking the pre-estimation effect into account. We present both analytical results and results based on the weighted bootstrap. We start by describing the setup and stating some technical conditions.

%We first present relatively simple results that consider individual points on the ROC curve, corresponding to fixed values of the cutoff $c$. We will then characterize the behavior of the entire sample ROC curve as a random function in the next section. 

\subsection{First stage estimation and technical assumptions}

The sample $\{(Y_i,X_i)\}_{i=1}^n$ now plays a dual role. First, it is used to construct an estimated parameter vector $\hat\beta=\hat\beta_n$. Typically, $\hat\beta$ consists of an intercept and slope coefficients from some type of regression of $Y$ on $X$ (e.g., linear, logit or probit). Second, the same sample is used to compute the predictive index values $G(X_i,\hat\beta)$, $i=1,\ldots, n$, and to construct the empirical ROC curve as described in Section \ref{subsec: sample ROC}. We impose the following high level condition on $\hat\beta$.

\begin{assumption}\label{assn: beta-est} (i) Let $\hat\beta$ be an M-estimator of $\beta$ so that
\begin{align*}
\hat{\beta}\equiv \arg\max_{\beta\in \mathcal{B}} \frac{1}{n} \sum_{i=1}^n q(Y_i,X_i,\beta).
\end{align*}
%where the parameter space $\mathcal{B}$ is a compact subset of $\subset\reals^p$. 
(ii) There is a point $\beta^*$ in the interior of the compact parameter space $\mathcal{B}\subset\reals^p$ so that
\begin{align}
\sqrt{n}(\hat{\beta}_n-\beta^*)=\frac{1}{\sqrt{n}}\sum_{i=1}^n\psi_\beta(Y_i,X_i,\beta^*)+o_p(1),
\label{eq: alpha-linear}
\end{align}
where $\psi_\beta: \reals^{1+k+p}\rightarrow \reals^p$ is a given function with $E[\psi_\beta(Y,X, \beta^*)]=0$ and $E\|\psi_\beta(Y,X, \beta^*)\|^{2+\epsilon}<\infty$ for some $\epsilon>0$. Furthermore, $\beta^*=\beta^\circ$ under Assumption~\ref{assn: correct spec}.
\end{assumption}

Assumption~\ref{assn: beta-est} states that $\hat\beta$ is an $M$-estimator with an asymptotically linear representation, implying that $\hat\beta$ is asymptotically normally distributed. The stated conditions do not require the first stage model to be correctly specified for $p(X)$; $\beta^*$ simply stands for the probability limit of $\hat\beta$,  i.e., the pseudo-true value of the parameter vector. We nevertheless assume that $\hat\beta$ is consistent for $\beta^\circ$ under correct specification (Assumption \ref{assn: correct spec}). The reason for allowing for misspecification is twofold. First, the first stage predictive model is often simply an approximation of $p(X)$, e.g., a linear regression. Second, as we will see, the estimation effect can depend on whether or not  the model is correctly specified.   

\begin{assumption}\label{assn: empirical process} 
The empirical processes
\[
(c,\beta)\mapsto\sqrt{n}(\hat \P_0-\P_0)1[G(X,\beta)\le c]\text{ and }(c,\beta)\mapsto\sqrt{n}(\hat \P_1-\P_1)1[G(X,\beta)\le c]
\]
are stochastically equicontinuous over $\reals\times \mathcal{B}$, where $\P_j$ denotes probability conditional on $Y=j$ and $\hat\P_j$ is the corresponding empirical measure in the $Y=j$ subsample. 
\end{assumption}

The stochastic equicontinuity requirement in Assumption \ref{assn: empirical process} limits the complexity of the model $G(X,\beta)$ and plays an important role in handling the estimation effect.
%fundamental role in the derivation of the uniform asymptotic results in Section~\ref{sec: uniform results}. 
It holds, for example, if $G(X,\beta)=X'\beta$ or $G(X,\beta)=G(X'\beta)$ with $G$ bounded (see the definition of a type I class in Andrews 1994). Apart from a small degree of added generality, we state stochastic equicontinuity as a high level condition to make it more transparent what is required for our results. 

The final assumption states the differentiability of $\TP$ and $\FP$ with respect to the components of $\beta$. Let $\nabla_\beta$ denote the corresponding gradient operator and $B^*(r)$ the open ball with radius $r>0$ centered on $\beta^*$. 
%Furthermore, $\mathcal{C}_j$ stands for the interior of the support of $G(X,\beta^*)$ conditional on  $Y=y$.  

%Let $\partial_{j}$ denote the partial derivative operator with respect to the $j$th component of $\beta$ with the corresponding gradient operator $\nabla_\beta=(\partial_1,\ldots,\partial_p)$. 

%The final set of assumptions state the existence of various derivatives of $G$, $\TP$ and $\FP$, and control their behavior. Let $\partial_{j}$ denote the partial derivative operator with respect to the $j$th component of $\beta$ and $\nabla_\beta$ the entire gradient. 
%Similarly, $\partial_{c}$ denotes partial derivative w.r.t\ $c$. 

\begin{assumption}\label{assn: gradient}
For any given cutoff $c$, the gradient vectors $\nabla_\beta\TP(c,\beta)$ and $\nabla_\beta\FP(c,\beta)$ exist and are continuous over $B^*(r)$ for some $r>0$.

\end{assumption}

As we will shortly see, the first stage estimation of $\beta$ affects the asymptotic distribution of the ROC curve through the derivatives presented in Assumption~\ref{assn: gradient}.

%\begin{assumption}\label{assn: gradient}
%(i) Let $\mathcal{C}=(a_1,b_1)$. $\TP(c,\beta)$ is twice continuously differentiable on $\mathcal{C}\times B^*(r)$ for some $r>0$  with $\sup_{(c,\beta)\in\mathcal{C}\times B^*(r)}\|\nabla_\beta\TP(c,\beta)\|\le M$ for some $M>0$. The same applies to $\FP$ with $\mathcal{C}=(a_0,b_0)$.
%
%(ii) For any given value of $X$, $G(X,\beta)$ is twice continuously differentiable w.r.t.\ $\beta$ on $B^*(r)$ for some $r>0$ with $\sup_{\beta\in B^*(r)}|\partial_{jj}G(X,\beta)|\le M$ almost surely for some $M>0$. 
%
%(iii) The conditional density of $G(X,\beta^*)$ given $\partial_jG(X,\beta^*)$ and $Y=y$ exists and is bounded uniformly by some $M>0$ for almost all values of $\partial_jG(X,\beta^*)$, $y=0,1$, and all $j$.   
%
%(iv) $E\big[|\partial_j G(X,\beta^*)|\,\big|\,Y=1\big]<\infty$ for all $j$.
%
%\end{assumption}

\subsection{Theoretical illustration of the estimation effect}\label{subsec: est eff simple theory}

Let $c$ be a given value of the cutoff; we want to conduct inference about the corresponding point $(\FP(c,\beta^*), \TP(c,\beta^*))$ on the limiting ROC curve. To isolate the effect of the first stage estimator $\hat\beta$ on the asymptotic distribution of $\widehat{\TP}(c,\hat\beta)$, we can write
\begin{eqnarray}
&&\sqrt{n}[\widehat{\TP}(c,\hat\beta)-\TP(c,\beta^*)]\nonumber\\
&&=\sqrt{n}[\widehat{\TP}(c,\hat\beta)-\TP(c,\hat\beta)]+\sqrt{n}[\TP(c,\hat\beta)-\TP(c,\beta^*)]\nonumber\\
&&=\sqrt{n}[\widehat{\TP}(c,\beta^*)-\TP(c,\beta^*)]+\sqrt{n}[\TP(c,\hat\beta)-\TP(c,\beta^*)]+o_p(1),\label{eq: est effect decomp 2nd eq}
\end{eqnarray}
where the second equality is due to the fact that the process $\sqrt{n}(\widehat{TP}-TP)$ is stochastically equicontinuous (Assumption~\ref{assn: empirical process}), implying
\[
\sqrt{n}[\widehat{\TP}(c,\hat\beta)-\TP(c,\hat\beta)]-\sqrt{n}[\widehat{\TP}(c,\beta^*)-\TP(c,\beta^*)]=o_p(1),
\]
given that $\hat\beta\rar_p\beta^*$. As $\beta^*$ is fixed and $Var[G(X,\beta^*)]>0$, the first term in equation (\ref{eq: est effect decomp 2nd eq}) has the asymptotic distribution given by (\ref{eq: \TP asy dist}) and the second term represents the effect of estimating $\beta^*$. Does this term have a non-negligible effect on the asymptotic distribution of $\widehat{\TP}(c,\hat\beta)$, and if yes, how do we characterize it?

To address these questions, we can use Assumption \ref{assn: gradient} to expand the second term in (\ref{eq: est effect decomp 2nd eq}) around $\beta^*$ to obtain  
\begin{eqnarray}
&&\sqrt{n}[\widehat{\TP}(c,\hat\beta)-\TP(c,\beta^*)]\nonumber\\
&&=\sqrt{n}[\widehat{\TP}(c,\beta^*)-\TP(c,\beta^*)]+\sqrt{n}\nabla_\beta \TP(c,\beta^*)(\hat\beta-\beta^*)+o_p(1).\label{est eff exp: eg}
\end{eqnarray}
%\begin{equation}\label{est eff exp: eg}
%    \sqrt{n}[\TP(c,\hat\beta)-\TP(c,\beta^\circ)]=\sqrt{n}\nabla_\beta \TP(c,\beta^\circ)(\hat\beta-\beta^\circ)+o_p(1).
%\end{equation}
Equation (\ref{est eff exp: eg}) shows that the estimation effect is negligible whenever $\nabla_\beta \TP(c,\beta^*)=0$. However, as Proposition \ref{prop: p(X) ROC opt}(ii) shows, this condition does not generally hold even if $G(X,\beta)$ is correctly specified (i.e., $\beta^*=\beta^\circ$), because $\TP(c,\beta^\circ)$ solves a \emph{constrained} (rather than unconstrained) optimization problem.\footnote{The first order conditions are $\nabla_\beta \TP(c,\beta^\circ)=\lambda\nabla_\beta \FP(c,\beta^\circ)$ for some scalar $\lambda$ and $\FP(c,\beta)=F^\circ_c$. The Lagrange multiplier $\lambda$ is generally non-zero, at least when $\TP(c,\beta^\circ)<1$.} Under misspecification Proposition \ref{prop: p(X) ROC opt}(ii) does not apply, but of course there is still no general reason for $\nabla_\beta \TP(c,\beta^*)$ to vanish. Therefore, in either case the asymptotic distribution of $\widehat{\TP}(c,\hat\beta)$ and $\widehat{\FP}(c,\hat\beta)$ will generally differ from that stated under (\ref{eq: \TP asy dist}) and (\ref{eq: \FP asy dist}) because $\sqrt{n}(\hat\beta-\beta^*)=O_p(1)$.\footnote{Proposition~\ref{prop: p(X) ROC opt}(i) implies that under correct specification one can conduct inference about the linear combination $(1-c)\pi\TP-c(1-\pi)\FP$ without the need to consider the pre-estimation effect. This is because the true value of $\beta$ maximizes this linear combination and hence the corresponding gradient driving the estimation effect vanishes. More generally, the estimation effect is negligible for a functional of the ROC curve if (i) the ROC curve based on $p(X)$ maximizes that functional and (ii) Assumption~\ref{assn: correct spec} holds.\label{fn: no est effect}} 

\subsection{Pointwise inference based on analytical results}
To describe the asymptotic distribution of $\widehat{\TP}(c,\hat\beta)$ in more detail, we can further expand the decomposition in (\ref{est eff exp: eg})  by substituting in the asymptotically linear (influence function) representation of the two terms. Using the definition of $\widehat{\TP}(c,\beta^*)$ and Assumption~\ref{assn: beta-est}, it is straightforward to verify that
\begin{align}
    \sqrt{n}[\widehat{\TP}(c,\hat\beta)-\TP(c,\beta^*)]&=\frac{1}{\sqrt{n}}\sum_{i=1}^n \Big\{\frac{Y_i}{\pi}\big[1(G(X,\beta^*)>c)-\TP(c,\beta^*)\big]\notag\\
    &~~+\nabla_\beta\TP(c,\beta^*)\psi_\beta(Y_i,X_i,\beta^*)\Big\}+o_p(1)\notag\\
    &\equiv \frac{1}{\sqrt{n}}\sum_{i=1}^n \psi_{\TP}(Y_i,X_i,c,\beta^*)+o_p(1),\label{eq: inf-TP}
\end{align}
where the definition of $\psi_{TP}$ is enclosed by the braces on the previous line.

Of course, $\widehat{\FP}(c,\hat\beta)$ has a corresponding asymptotically linear representation with influence function 
\[
\psi_{\FP}(Y_i,X_i,c,\beta^*)=\frac{1-Y_i}{1-\pi}\big[1(G(X_i,\beta^*)>c)-\FP(c,\beta^*)\big]+\nabla_\beta\FP(c,\beta^*)\psi_\beta(Y_i,X_i,\beta^*).
\]
Stacking the influence functions as $$\psi(Y_i,X_i,c,\beta^*)=[\psi_{\TP}(Y_i,X_i,c,\beta^*),\psi_{\FP}(Y_i,X_i,c,\beta^*)]'$$ 
and applying the multivariate CLT gives the asymptotic joint distribution of an individual point $(\widehat{\TP}(c,\hat\beta),\widehat{\FP}(c,\hat\beta))$ on the sample ROC curve.   

\begin{proposition}\label{prop: est effect gen}
Suppose that Assumptions \ref{assn: iid} to \ref{assn: gradient} are satisfied. Then
\begin{equation}\label{eq: TF joint dist}
    \sqrt{n}
    \begin{pmatrix}
    \widehat{\TP}(c,\hat\beta)-\TP(c,\beta^*)\\
    \widehat{\FP}(c,\hat\beta)-\FP(c,\beta^*)\\
    \end{pmatrix}
        =\frac{1}{\sqrt{n}}\sum_{i=1}^n \psi(Y_i,X_i,c,\beta^*)+o_p(1)
    \rightarrow_d N[0,E(\psi\psi')]
\end{equation}
for cutoffs $c$ for which $E[\psi^2_{\TP}(Y_i,X_i,c,\beta^*)]>0$ and $E[\psi^2_{\FP}(Y_i,X_i,c,\beta^*)]>0$.
%$\P[G(X,\beta^*)>c|Y=y]\in (0,1)$, $y=0,1$. 
\end{proposition}

\paragraph{Remarks}
\begin{enumerate}
    
    \item Using Proposition~\ref{prop: est effect gen}, it is easy to obtain the asymptotic distribution of any linear combination $a\widehat{\TP}(c,\hat\beta)+b\widehat{\FP}(c,\hat\beta)$. 
    %In particular, for $\pi=(a,b)'$,
    %\[
    %\sqrt{n}[a\widehat{\TP}(c,\hat\beta)+b\widehat{\FP}(c,\hat\beta)-(a\TP+b\FP)]\rar_d N\big(0,\pi'ME[\psi\psi']M'\pi\big)
    %\]
    %For example, inference about $\TP(c,\beta^\circ)-\FP(c,\beta^\circ)$ is helpful to decide whether applying the given cutoff to the index gives better predictions than a random guess.\footnote{For any given value of \FP one can always achieve the same level of \TP simply by flipping an unbalanced coin with head probability \FP and predicting $Y=1$ if that event occurs.}
    
    \item Proposition \ref{prop: est effect gen} is a ``pointwise'' result in the sense that the cutoff $c$ is assumed to be fixed. It is straightforward to generalize the setup so that one can make joint inference about points that are associated with a finite number of different cutoffs. One can simply stack the values of the influence function $\psi$ evaluated at these cutoffs and a result analogous to (\ref{eq: TF joint dist}) will continue to hold.
    
    \item The variance condition $E[\psi^2_{TP}]>0$ will generally hold for interior points $\TP(c,\beta^*)\in (0,1)$ but fail for $\TP(c,\beta^*)\in \{0,1\}$. The same is true for $\FP$.
    
    %\item To use Proposition~\ref{prop: est effect logit} in practice, one must of course estimate $\psi$ and $M$. As discussed above, the gradient vectors can be estimated by (\ref{1 step grad est}) and substituted in the definition of $M$ to obtain $\hat M$. To construct $\hat\psi$, one can replace unknown parameters by their empirical counterparts and expectations 
    %in the definition of $A_\beta$ by the corresponding sample average. Then one estimates the variance-covariance matrix of the limit distribution $(\ref{eq: dist TF est effect})$ by $\hat M(n^{-1}\sum_{i=1}^n \hat\psi_i\hat\psi_i')\hat M'$.
            
    %\item If the first stage is a logit regression based on a linear index and estimated by maximum likelihood, then Assumption \ref{assn: stoch equicont}(i) and (ii) are generally satisfied subject to some auxiliary technical conditions. %The main reason why we state the assumption anyway is to make it clearer what drives the results and 
    %Nevertheless, stating this high level assumption (rather than technical primitives) makes for a simpler exposition and allows for logit regression to be replaced by other (potentially more complicated) first stage estimation procedures.
    
\end{enumerate}
    
We supplement Proposition \ref{prop: est effect gen} by some results that reveal the structure of $\nabla_\beta \TP$ and $\nabla_\beta \FP$ and facilitate their estimation. Let $\partial_{j}$ denote the partial derivative operator with respect to the $j$th component of $\beta$. 

\begin{assumption}\label{assn: gradient2}
(i) $G(X,\beta)$ is twice continuously differentiable (a.s.) w.r.t.\ $\beta$ on $B^*(r)$ for some $r>0$ with $\sup_{\beta\in B^*(r)}|\partial_{jj}G(X,\beta)|\le M$ (a.s.) for some $M>0$. 

(ii) The conditional density of $G(X,\beta^*)$ given $Y=0,1$ exists. The conditional density of $G(X,\beta^*)$ given $\partial_jG(X,\beta^*)$ and $Y=y$ also exists and is bounded uniformly by some $M>0$ for almost all values of $\partial_jG(X,\beta^*)$, $y=0,1$, and all $j$.   

(iii) $E\big[|\partial_j G(X,\beta^*)|\,\big|\,Y=1\big]<\infty$ for all $j$.
\end{assumption}

\begin{proposition}\label{prop: gradient \TP} Suppose that Assumptions~\ref{assn: gradient} and \ref{assn: gradient2} hold. Then:
\begin{equation}\label{eq: gradient \TP}
\nabla_\beta \TP(c,\beta^*)=E\Big[\nabla G_\beta(X,\beta^*)\Big|\,G(X,\beta^*)=c, Y=1 \Big]f^*_1(c),
\end{equation}
where $f^*_1(c)$ is the conditional density of $G(X,\beta^*)$ given $Y=1$. If, in addition, Assumption~\ref{assn: correct spec} is satisfied ($\beta^*=\beta^\circ$), then the expectation in equation (\ref{eq: gradient \TP}) does not need to be conditioned on $Y=1$. The formula for $\nabla_\beta \FP(c,\beta^*)$ is analogous; it conditions on $Y=0$ throughout.

\end{proposition}

Finally, we specialize Propositions~\ref{prop: est effect gen} and \ref{prop: gradient \TP} by imposing a logit first stage.

\begin{assumption}\label{assn: logit}
Suppose that the first stage estimation consists of a logit regression of $Y$ on $X$ and a constant so that $G(X,\hat\beta)=\Lambda(\tilde X'\hat\beta)$, where $\Lambda(\cdot)$ is the logistic c.d.f., $\tilde X=(1, X')'$ and $\hat\beta$ is the maximum likelihood estimator. 
\end{assumption}
\begin{proposition}\label{prop: logit} 
Suppose that Assumption~\ref{assn: logit} is satisfied. Then: 
\begin{itemize}
    \item [(a)] $\psi_\beta(Y_i,X_i,\beta)=A_\beta^{-1}X_i[Y_i-\Lambda(\tilde X_i'\beta)]$, where $A_\beta=E\{\Lambda(\tilde X'\beta)[1-\Lambda(\tilde X'\beta)]\tilde X\tilde X'\}$.
    \item [(b)] The components of $\nabla_\beta \TP(c,\beta^*)$ are given by:
\begin{equation}\label{eq: \TP grad sp}
    c(1-c)E\big[X_j\big|\,\Lambda(\tilde X'\beta^*)=c, Y=1 \big]f_1^*(c),\; j=0,1,\ldots,d,
\end{equation}
where $X_0\equiv 1$, $X_j$, $j=1,\ldots,d$ is the $j$th component of $X$, and $f_1^*(c)$ is the conditional density of $\Lambda(\tilde X'\beta^*)$ given $Y=1$.
%(ii) If, in addition, $X\sim N(\mu,\Sigma)$, then the components of $\nabla_\beta \TP(c,\beta^\circ)$ are given by:
%\begin{equation}\label{eq: \TP grad pm}
%    \frac{c}{\pi}\big[\theta_{0j}+\theta_{1j}\Lambda^{-1}(c)\big]
%    \frac{1}{\sqrt{\beta_1'\Sigma\beta_1}}\phi\left(\frac{\Lambda^{-1}(c)-\beta_0-\mu'\beta_1}{\beta_1'\Sigma\beta_1}\right),\; j=0,1,\ldots,d,
%\end{equation}
%where $\theta_{0j}$ and $\theta_{1j}$ are the coefficients from the linear projection of $X_j$ on a constant and $\tilde X'\beta^\circ$, $\phi$ is the standard normal p.d.f.\ and $\beta^\circ=(\beta_0,\beta_1')'$ with $\beta_0$ being a scalar.  
\end{itemize}

\end{proposition}

\paragraph{Remarks:}

\begin{enumerate}
    
    \item The proofs of Propositions \ref{prop: gradient \TP} and \ref{prop: logit} are presented in Appendix B.
    
    \item The existence of $f_1^*(c)$ requires that $X$ has a continuous component and the corresponding coefficient in $\beta^*$ is nonzero. This rules out $X$ and $Y$ being independent. 
    
    %A first stage logit regression estimated by maximum likelihood generally satisfies the technical conditions in Assumption~\ref{assn: gradient} provided that $X$ has a continuous component with a nonzero coefficient.
    
    \item The expression for $\psi_\beta$ follows from formulas (12.16), (15.18) and (15.19) in Wooldridge (2002) when specialized to the logit case.
    
    \item One can estimate the unknown quantities in (\ref{eq: \TP grad sp}) nonparametrically to obtain a semiparametric estimator for $\nabla_\beta \TP(c,\beta^*)$. More precisely, expression (\ref{eq: \TP grad sp}) may actually be estimated in a single step as  
    \begin{equation}\label{1 step grad est}
    c(1-c)\frac{1}{n_1h}\sum_{i: Y_i=1}X_{ji}K\left(\frac{\Lambda(\tilde X_i'\hat\beta)-c}{h}\right),
    \end{equation}
    where $K(\cdot)$ is a kernel function and $h$ is a bandwidth that may be chosen according to Silverman's rule of thumb. 
    
    \item Alternatively, if correct specification is assumed in the first stage ($\beta^*=\beta^\circ$), then one can estimate $E[X_j|\,\Lambda=c, Y=1]=E[X_j|\,\Lambda=c]$
    %$E\big[X_j\big|\,\Lambda(\tilde X'\beta^\circ)=c\big]$
    by a kernel regression on the \emph{full sample} and $f^*_1(c)$ by a kernel estimator on the $Y=1$ subsample. %The derivative (\ref{eq: \TP grad sp}) is then the product of $c(1-c)$ and these two estimates. 

\end{enumerate}

\subsection{Pontwise inference based on the weighted bootstrap}\label{subsec: weighted}

Here we provide an alternative method for making pointwise inference about the ROC curve by utilizing the weighted bootstrap for M-estimators proposed by Ma and Kosorok (2005). The main advantage of this approach is that it sidesteps the estimation of the gradient vectors $\nabla_\beta \TP(c,\beta^*)$ and $\nabla_\beta \FP(c,\beta^*)$. Furthermore, the method is similar to the simulation-based procedure that we propose for functional inference in Section \ref{sec: uniform results}.

The weighted bootstrap employs a sequence of (pseudo) random variables as multipliers to simulate the sampling variation of an estimator. 

\begin{assumption}\label{assn: weighted bootstrap W}
Let $\{W_i\}_{i=1}^n$ be a sequence of i.i.d.\ (pseudo) random variables,  independent of the sample path $\{(Y_i,X_i)\}_{i=1}^n$, with $E(W_i)=1$ and $Var(W_i)=1$. 
\end{assumption}

We first define the weighted bootstrap version of the first stage estimator of $\beta$:  
\begin{align*}
\hat{\beta}^w= \arg\max_{\beta\in \mathcal{B}} \frac{1}{n} \sum_{i=1}^n W_i \cdot q(Y_i,X_i,\beta).
\end{align*}

Given $\hat{\beta}^w$, the weighted bootstrap estimators of $\TP(c,\beta)$ and $\FP(c,\beta)$ are defined as 
\begin{eqnarray*}
\widehat{\TP}^w(c,\beta)&=&\frac{1}{\sum_{i=1}^n W_i\cdot Y_i} \sum_{i=1}^n W_i \cdot 1[G(X_i,\beta)>c, Y_i=1]\\
\widehat{\FP}^w(c,\beta)&=&\frac{1}{\sum_{i=1}^n W_i \cdot(1-Y_i)} \sum_{i=1}^n W_i \cdot 1[G(X_i,\beta)>c, Y_i=0].
\end{eqnarray*}

\begin{assumption}\label{assn: beta-est-weighted boot} Assume that 
\begin{align}
\sqrt{n}(\hat{\beta}^w-\beta^*)=\frac{1}{\sqrt{n}}\sum_{i=1}^nW_i \cdot \psi_\beta(Y_i,X_i,\beta^*)+o_p(1),
\end{align}
where $\beta^*$ and $\psi_\beta(Y_i,X_i,\beta^*)$ are given in Assumption \ref{assn: beta-est}.
\end{assumption}

Assumption~\ref{assn: beta-est-weighted boot} ensures that the weighted bootstrap is valid for the first stage estimator, i.e.,  conditional on the data, $\sqrt{n}(\hat{\beta}^w-\hat{\beta})$ has the same limiting distribution as $\sqrt{n}(\hat\beta-\beta^*)$ unconditionally.

Furthermore, by Theorem 2 of Ma and Kosorok (2005), the validity of the weighted bootstrap for $\widehat{\TP}^w(c,\hat\beta^w)$
%, in the sense that conditional on the data, $\sqrt{n}[\widehat{\TP}^w(c,\hat\beta^w)-\widehat{\TP}(c,\hat\beta)]$ has the same limiting distribution as $\sqrt{n}[\widehat{\TP}(c,\hat\beta)-\TP(c,\beta^*)]$ unconditionally, 
follows from showing that (i) $\widehat{\TP}$, $\widehat{\FP}$, $\widehat{\TP}^w$ and $\widehat{\FP}^w$ can be represented as M-estimators and (ii) that these estimators are $\sqrt{n}$-consistent and asymptotically linear.

Item (i) is verified by noting that 
\begin{eqnarray*}
\widehat{\TP}(c,\hat{\beta})&=&\arg\min_{t\in \reals }\frac{1}{n}\sum_{i=1}^n Y_i\cdot  
\big(1[G(X_i,\hat{\beta})>c]-t \big)^2\\
\widehat{\TP}^w(c,\hat{\beta}^w)&=&\arg\min_{t\in \reals}\frac{1}{n}\sum_{i=1}^n W_i\cdot Y_i\cdot  
\big(1[G(X_i,\hat{\beta}^w)>c]-t \big)^2,
\end{eqnarray*} 
and similarly for $\widehat{\FP}$ and $\widehat{\FP}^w$. As for item (ii), Proposition \ref{prop: est effect gen} establishes the asymptotically linear representation of $(\widehat{\TP}(c,\hat\beta), \widehat{\FP}(c,\hat\beta))$; essentially the same argument also yields
\begin{equation*}
    \sqrt{n}
    \begin{pmatrix}
    \widehat{\TP}^w(c,\hat\beta^w)-\TP(c,\beta^*)\\
    \widehat{\FP}^w(c,\hat\beta^w)-\FP(c,\beta^*)\\
    \end{pmatrix}
        =\frac{1}{\sqrt{n}}\sum_{i=1}^n W_i\cdot \psi(Y_i,X_i,c,\beta^*)+o_p(1)\label{eq: weight B 1}.
\end{equation*}

Thus, we obtain the following result.
\begin{proposition}\label{prop: est effect gen-bootstrap}
Suppose that Assumptions \ref{assn: iid}-\ref{assn: gradient}, \ref{assn: weighted bootstrap W} and 
\ref{assn: beta-est-weighted boot} are satisfied. Then, conditional
on the sample path of the data, 
\begin{equation*}
    \sqrt{n}
    \begin{pmatrix}
    \widehat{\TP}^w(c,\hat\beta^w)-\widehat{\TP}(c,\hat{\beta})\\
    \widehat{\FP}^w(c,\hat\beta^w)-\widehat{\FP}(c,\hat{\beta})\\
    \end{pmatrix}
    =\frac{1}{\sqrt{n}}\sum_{i=1}^n (W_i-1)\cdot \psi(Y_i,X_i,c,\beta^*)
    \rightarrow_d N[0,E(\psi\psi')]
\end{equation*}
with probability approaching one for cutoffs $c$ such that $E[\psi^2_{\TP}]>0$ and $E[\psi^2_{\FP}]>0$. 
\end{proposition}

\paragraph{Remarks}

\begin{enumerate}
    
    \item In applications we suggest letting the weights $W_i$ take the values 0 and 2 with equal probability. The main reason is that with non-negative weights the weighted objective function remains concave if the $q(Y_i,X_i,\beta)$ is concave in $\beta$. This makes it computationally easier to obtain $\hat\beta^w$.   
    
    \item The weighted bootstrap estimator of the asymptotic variance-covariance matrix $\Psi(c)\equiv E(\psi\psi')$ can be constructed as follows. With a minor abuse of notation, let $\widehat R^w(c)=(\widehat{\TP}^w(c,\hat\beta^w), \widehat{\FP}^w(c,\hat\beta^w))'$ denote the ROC estimate from the $w$th bootstrap cycle, $w=1,\ldots, \mathcal{W}$. Then one can estimate $\Psi(c)$ by    
    \begin{align*}
    &\widehat\Psi_\mathcal{W}(c)=\frac{n}{\mathcal{W}}\sum_{w=1}^\mathcal{W} \big(\widehat{R}^w(c)-\overline{\widehat{R}}^w(c)\big)\big(\widehat{R}^w(c)-\overline{\widehat{R}}^w(c)\big)',\text{ where}\\
   &\overline{\widehat{R}}^w(c)=\frac{1}{\mathcal{W}}\sum_{w=1}^\mathcal{W}\widehat{R}^w(c).
\end{align*}
We have that conditional on sample path with probability approaching one,
\begin{align*}
&\widehat{\Psi}_{\mathcal{W}}(c)\stackrel{p}{\rightarrow}_w
\frac{1}{n}\sum_{i=1}^n \psi(Y_i,X_i,c,\beta^*)\psi(Y_i,X_i,c,\beta^*)'+o_p(1),
\end{align*}
where $\stackrel{p}{\rightarrow}_w$ denotes probability limit under the law of the $W_i$'s. It follows that
\begin{align*}
\lim_{\mathcal{W}\rightarrow \infty} \widehat{\Psi}_\mathcal{W}(c)\stackrel{p}{\rightarrow} {\Psi}(c).
\end{align*}

\end{enumerate}

\section{In-sample inference: uniform asymptotics}\label{sec: uniform results}

%\subsection{Revisiting the definition of the ROC curve} 
 
%Our goal is to study the sample ROC curve based on a pre-estimated index $G(X,\hat\beta)$ as a random \emph{function} over the interval $[0,1]$, representing the range of possible $\FP$ rates. 

To derive uniform results, we first express the ROC curve explicitly as a function over the interval [0,1]. Let the inverse of the \emph{decreasing} function $c\mapsto FP(c,\beta)$ be defined as
\[
\FP^{-1}_{\beta}(t)=\inf\{c: FP(c,\beta)\le t\},\; t\in [0,1].
\]
The more compact notation on the l.h.s.\ emphasizes that the inverse is taken with respect to the cutoff $c$ for a fixed value of $\beta$. Thus, is $\FP^{-1}_{\beta}(t)$ as the ``first'' (smallest) cutoff value at which the false positive rate is equal to $t$ or falls below $t$.\footnote{Of course, if $\FP(c,\beta)$ is strictly decreasing and continuous in $c$, then $\FP^{-1}_{\beta}(t)$ is the unique solution to the equation $\FP(c,\beta)=t$.} 
%We will therefore use the further abbreviation $c_t:=\FP^{-1}_{\beta}(t)$. 
Because $1-FP(c,\beta)$ is the c.d.f.\ of the conditional distribution of $G(X,\beta)$ given $Y=0$, an equivalent interpretation of $\FP^{-1}_{\beta}(t)$ is that it is the $(1-t)$-quantile of this distribution.

%While $\FP^{-1}_{\beta}(t)$ is not a proper quantile function, it has an analogous interpretation: $\FP^{-1}_{\beta}(t)$ returns the ``first'' (smallest) value of $c$ where $\FP(c,\beta)$ is equal to $t$ or dips \emph{below} $t$.  Of course, if $\FP(c,\beta)$ is strictly decreasing and continuous in $c$, then the equation $\FP(c,\beta)=t$ has a unique solution for all $t$. Thus, for any $t\in [0,1]$, representing a given false positive rate, one can simply think of $\FP^{-1}_{\beta}(t)$ as the cutoff value that gives rise to that false positive rate.

We can now represent the ROC curve as a function that returns the true positive rate associated with given false positive rate $t$:
\begin{equation}\label{def: ROC pop explicit}
R(t,\beta)=\TP\big(\FP^{-1}_{\beta}(t),\beta\big),\quad t\in[0,1].
%=1-FN\big(\TN^{-1}_{\beta}(1-t),\beta\big),\quad t\in[0,1].
\end{equation}
%The second expression is less intuitive but it has the advantage that $\FN(\cdot,\beta)$ is a proper cdf and $Q_{TN}(\cdot,\beta)$ is a proper quantile function. This makes it more straightforward to apply results from the vast statistical literature on the estimation of these functions. 
For a given parameter value $\beta$, the sample ROC curve is defined by replacing $\TP(\cdot,\beta)$ and $\FP^{-1}_\beta(\cdot)$ by sample analogs: $\widehat R(t,\beta)=\widehat{\TP}\big(\widehat{\FP}^{-1}_{\beta}(t),\beta\big),\;t\in[0,1]$.
%\begin{equation*}\label{def: ROC smpl explicit}
%\widehat R(t,\beta)=\widehat{\TP}\big(\widehat{\FP}^{-1}_{\beta}(t),\beta\big),\quad t\in[0,1].
%=1-\widehat{\FN}\big(\widehat{\TN}^{-1}_{\beta}(1-t),\beta\big),\quad t\in[0,1],
%\end{equation*}
%where $\widehat{\FP}^{-1}_{\beta}(t)=\inf\{c: \widehat{\FP}(c,\beta)\le t\}$.
%$$\widehat{TP}(c,\beta)=\frac{1}{n_1}\sum_{i: Y_i=1} 1[G(X_i,\beta)>c],\quad \widehat{ \FP}(c,\beta)=\frac{1}{n_0}\sum_{i: Y_i=0} 1[G(X_i,\beta)> c],$$
%and $\widehat{\FP}^{-1}_{\beta}(t)=\inf\{c: \widehat{\FP}(c,\beta)\le t\}$.%\footnote{One can also define the ROC curve in terms of the true and false negative rates, i.e., $\TN(\beta,c)=\P[G(X,\beta)\le c|Y=0]$ and $\FN(\beta,c)=\P[G(X,\beta)\le c|Y=0]$. In this case $R(t,\beta)=1-FN\big(\TN^{-1}_{\beta}(1-t),\beta\big)$. The slight advantage of this formulation is that $\FN$ is cdf and $\TN^{-1}$ is a proper quantile function.}  

\subsection{Additional technical assumptions for uniform inference}

Our goal is to characterize the statistical behavior of the random function $t\mapsto \hat R(t,\hat\beta)$ over the interval $[0,1]$. This requires some additional assumptions.  

\begin{assumption}\label{assn: cond densities} 
(i) The conditional distribution of $G(X,\beta^*)$ given $Y=0$ has compact support $[a_0,b_0]$ and probability density function $f_0^*(c)$ that is continuous (and hence bounded) over $[a_0,b_0]$ and satisfies $\inf\{f_0^*(c): c\in[a_0,b_0]\}\ge\delta$ for some $\delta>0$.  

(ii) The conditional distribution of $G(X,\beta^*)$ given $Y=1$ has compact support $[a_1,b_1]$ and a probability density function $f_1^*(c)$.
%that is continuous over the interior of $[a_1,b_1]$.

%(i) The conditional distribution of $G(X,\beta^*)$ given $Y=y$ is absolutely continuous with density function $f^*_y(c)$, $y=0,1$.

%(ii) The support of $f_0^*$ is a compact interval $[a_0,b_0]$. $f^*_0$ is continuous and bounded away from zero over $[a_0,b_0]$, i.e., $\inf_{c\in[a_0,b_0]} f_0(c)\ge\delta$ for some $\delta_0>0$. $f_1^*$ is continuous over the interior of its support.
 
(iii) There exits a subinterval $[c_{0,L},c_{0,U}]\subseteq [a_0,b_0]$ such that $f^*_1(c)/f^*_0(c)$ is continuous (and hence bounded) over $[c_{0,L},c_{0,U}]$ and satisfies $\inf\{f_1^*(c)/f_0^*(c): c\in[c_{0,L},c_{0,U}]\}\ge\delta$ for some $\delta>0$.
\end{assumption}

Assumption~\ref{assn: cond densities} merits careful discussion. An immediate practical implication of part (i) is that the limiting model $G(X,\beta^*)$ must depend on at least one continuous predictor in a nontrivial way. For instance, if the model is based on a linear index, this rules out $X$ being completely independent of $Y$; see Remark~1 after Proposition~\ref{prop: logit}. Part (iii) implies that $supp(f_0^*)$ and $supp(f_1^*)$ overlap, ensuring that the classification problem is nontrivial. Nevertheless, the overlap does not need to be complete; we allow for applications in which extreme values of the index are associated exclusively with one of the two outcomes.

From a technical standpoint, the main purpose of Assumption~\ref{assn: cond densities} is to facilitate uniform inference by controlling the behavior of the likelihood ratio $f_1^*/f_0^*$. In particular, $f_1^*/f_0^*$ is required to be bounded and bounded away from zero on an interval $[c_{0,L},c_{0,U}]$. Our uniform influence function representation result for $\hat R(t,\hat\beta)$ holds only for quantiles $t$ satisfying $\FP^{-1}_{\beta^*}(t)\in [c_{0,L},c_{0,U}]$ or, equivalently, for $t\in [\FP(c_{0,U},\beta^*), \FP(c_{0,L},\beta^*)]$. While this representation depends on $f_1^*$ and $f_0^*$ only through $f_1^*/f_0^*$, the derivation of the result relies on the additional condition that $f_0^*$ is bounded away from zero (Assumption~\ref{assn: cond densities}(i)). This may seem overly restrictive at first glance---for example, if $G(X,\beta^*)=X'\beta^*$, then predictors with unbounded support are ruled out. Furthermore, it is easy to see that even if all components of $X$ have densities bounded away from zero, their linear combinations will generally not share this property.\footnote{For example, consider the sum of two independent uniform [-0.5,0.5] random variables. The resulting density is $(1-|x|)1_{[-1,1]}(x)$, which tends to zero as $x$ approaches $-1$ or $1$. 
%Or consider the central limit theorem.
} 
However, one can always find a monotone increasing transformation $\Phi(\cdot)$ such that the density of $\Phi[G(X,\beta^*)]$ conditional on $Y=0$ is bounded away from zero, e.g., one can use the probability integral transform to arrive at a uniform[0,1] density. At the same time, such a transformation leaves the ROC curve as well as the range of the likelihood ratio $f_1^*/f_0^*$ unchanged. Thus, the last part of Assumption \ref{assn: cond densities}(i) is simply a theoretical normalization that does not need to be imposed on the data in practice (see Figure~1 for an illustration).

\begin{figure}[!t]\label{fig:densities}
    \begin{center}
    \includegraphics[scale=0.95]{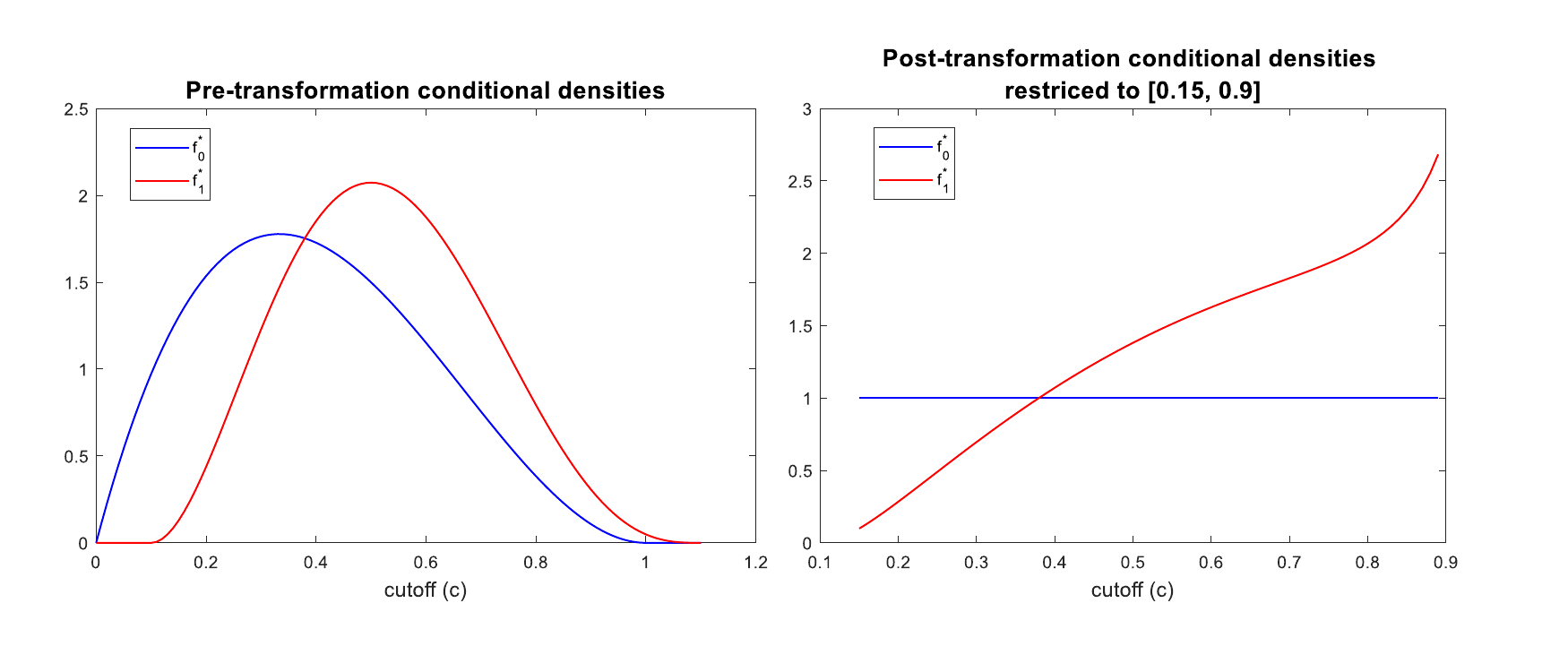}
    \caption{ {\footnotesize On the left panel, $f_0^*$, the blue curve, is the $\beta(2,3)$ pdf so that $[a_0,b_0]=[0,1]$ and $f^*_0$ is not bounded away from zero. $f_1^*$, the red curve, is $\beta(3,4)+0.1$ so that $[a_1,b_1]=[0.1,1.1]$. The likelihood ratio is zero below 0.1 and becomes unbounded just below 1. On the right panel, we apply the transformation $\Phi=\,$cdf of $\beta(2,3)$. $f^*_0$ is now the uniform[0,1] density so that $f_1^*$ transforms into the likelihood ratio. Assumption~\ref{assn: cond densities}(iii) is satisfied over, say, $[c_{0,L},c_{0,U}]=[0.15,0.9]$ so that uniform inference about $R(t,\beta^*)$ is possible over $[\FP(c_{0,U},\beta^*), \FP(c_{0,L},\beta^*)]\approx [0.004,0.89]$. }}
    \end{center}
\end{figure}

Of course, Assumption~\ref{assn: cond densities} allows for scenarios in which $[\FP(c_{0,U},\beta^*), \FP(c_{0,L},\beta^*)]=[0,1]$, i.e., uniform inference is possible along the entire ROC curve. This is the case, for example, if the ``propensity score'' function $P(Y=1|X=x)$ takes values from an interval $[\delta, 1-\delta]$ for some $0<\delta<1/2$, which implies $supp(f_0^*)=supp(f_1^*)$ and that (iii) holds with $[c_{0,L},c_{0,U}]=[a_0,b_0]$.\footnote{To see this, let $f_x(x)$ denote the density function of $X$. Note that 
\begin{align*}
&f_0(c)=\frac {\int_{G(x)=c}  (1-p(x)) f_x(x) dx}{\int_{G(x)=c}  (1-p(x)) dx }
\quad\text{ and }\quad f_1(c)=\frac {\int_{G(x)=c}  p(x) f_x(x) dx}{\int_{G(x)=c}  p(x) dx }.
\end{align*}
It follows that
\begin{align*}
\frac{f_0(c)}{f_1(c)}=\frac {\int_{G(x)=c}  (1-p(x)) f_x(x) dx}{\int_{G(x)=c}  p(x) f_x(x) dx} 
\frac{\int_{G(x)=c}  p(x) dx } {\int_{G(x)=c}  (1-p(x)) dx },
\end{align*}
which is bounded below by $ \delta^2/(1-\delta)^2$ and bounded above by 
$ (1-\delta)^2/\delta^2$. } More generally, Assumption~\ref{assn: cond densities}(iii) allows $f^*_1(c)/f^*_0(c)$ to reach zero or explode for cutoffs $c$ outside the range $[c_{0,L}, c_{1,L}]$. For example, the likelihood ratio vanishes as $c$ approaches $a_0$ from above whenever $a_0<a_1$. In this case the lowest index values imply $Y=0$ and the ROC curve reaches the top of the unit square for some $\FP$ rate below unity. Similarly, $f^*_1(c)/f^*_0(c)$ may become unbounded as $c$ approaches $b_0$ from below. This can easily happen when $b_0<b_1$, i.e., the largest index values are associated exclusively with the $Y=1$ outcome. In this case the ROC curve has a positive vertical intercept at $\FP=0$. Again, see Figure~1 for an example. %[\textcolor{red}{picture?}]    

The next assumption is a strengthening of Assumption~\ref{assn: gradient}. 
%and ensures the uniform continuity of the gradient vectors $\nabla_\beta TP(c,\beta)$ and $\nabla_\beta FP(c,\beta)$ over $[a_j,b_j]\times B^*(r)$. 
These stricter conditions on the gradient vectors $\nabla_\beta TP(c,\beta)$ and $\nabla_\beta FP(c,\beta)$ also play a key role in establishing a uniform influence function representation for the sample ROC curve. Recall that $B^*(r)$ denote the open ball with radius $r>0$ centered on $\beta^*$.

\begin{assumption}\label{assn: gradient unif}
Let $\mathcal{C}=[a_1,b_1]$. $\nabla_\beta\TP(c,\beta)$ exits and is continuous over $\mathcal{C}\times B^*(r)$ for some $r>0$  with $\sup_{(c,\beta)\in\mathcal{C}\times B^*(r)}\|\nabla_\beta\TP(c,\beta)\|\le M$ for some $M>0$. The same applies to $\nabla_\beta\FP(c,\beta)$ with $\mathcal{C}=[a_0,b_0]$.
\end{assumption}

\subsection{Functional limit results}

Letting $c^*_t={\FP}_{\beta^*}^{-1}(t)\in[a_0,b_0]$ and $\hat c_t= \widehat{\FP}_{\hat\beta}^{-1}(t)\in[a_0,b_0]$, we start from a decomposition of $\sqrt{n}[\widehat{\TP}(\hat c_t,\hat\beta)-\TP(c^*_t,\beta^*)]$ similar to (\ref{eq: est effect decomp 2nd eq}). There are two added layers of difficulty. First, functional results require uniform approximations to these terms as $t$ varies over the $[0,1]$ interval. Second, instead of being fixed, the cutoff is now estimated for any given value of $t$. The sampling variation in $\hat c_t$ contributes another non-trivial term to the asymptotic distribution. 

We express the centered and scaled empirical ROC curve as the sum of three terms:
\begin{multline}
\sqrt{n}[\widehat{R}(t,\hat\beta)-R(t,\beta^*)]=\sqrt{n}[\widehat{\TP}(\hat c_t,\hat\beta)-\TP(c^*_t,\beta^*)]\\
=\sqrt{n}[\widehat{\TP}(\hat c_t,\hat\beta)-\TP(\hat c_t,\hat\beta)]+\sqrt{n}[\TP(\hat c_t,\hat\beta)-\TP(\hat c_t,\beta^*)]\\
+\sqrt{n}[\TP(\hat c_t,\beta^*)-\TP(c^*_t,\beta^*)]\label{eq: func decomp}
\end{multline}

The first term in equation (\ref{eq: func decomp}) can be expanded similarly to the second equality in (\ref{eq: est effect decomp 2nd eq}):
\begin{equation}\label{eq: R1}
\sqrt{n}[\widehat{\TP}(\hat c_t,\hat\beta)-\TP(\hat c_t,\hat\beta)]=\sqrt{n}[\widehat{\TP}(c^*_t,\beta^*)-\TP(c^*_t,\beta^*)]+R_{1n}(t), 
\end{equation}
where $\sup_{t\in[0,1]}|R_{1n}(t)|=o_p(1)$. The uniform convergence of the remainder term is a consequence of the stochastic equicontinuity of the process $(c,\beta)\mapsto \sqrt{n}(\widehat{\TP}(c,\beta)-\TP(c,\beta))$, stated directly in Assumption~\ref{assn: empirical process}, coupled with the  
fact that $\hat\beta\rar_p\beta^*$ (Assumption~\ref{assn: beta-est}) and $\sup_{t\in[0,1]}|\hat c_t-c^*_t|\rar_p 0$ (Lemma~\ref{lm: hatct to cstart} in Appendix~A). This last result makes use of Assumption~\ref{assn: cond densities}(i), which requires that the density $f_0^*$ be bounded away from zero on its compact support. %Lemma \ref{lm: unif R} in Appendix~A formalizes the preceding arguments.  

The second term in equation (\ref{eq: func decomp}) is due to the estimation of $\beta$ and is again handled by a standard mean value expansion:
\begin{equation}\label{eq: R2}
\sqrt{n}[\TP(\hat c_t,\hat\beta)-\TP(\hat c_t,\beta^*)]=\nabla_\beta\TP(c^*_t,\beta^*)\sqrt{n}(\hat\beta-\beta^*)+R_{2n}(t),
\end{equation}
where $\sup_{t\in[0,1]}|R_{2n}(t)|=o_p(1)$. The uniformity of the approximation is ensured by Assumption~\ref{assn: gradient unif}, which implies that $\nabla_\beta\TP(c,\beta^*)$ is uniformly continuous. 
%; see again Lemma \ref{lm: unif R} in Appendix~A. 

Finally, the third term in (\ref{eq: func decomp}) arises because of the need to estimate the cutoff value associated with a given false positive rate $t$; it therefore does not arise in the fixed-cutoff setting. Starting with a mean value expansion of $\TP(\hat c_t,\beta^*)$ around $c^*_t$, one can write
\begin{eqnarray}
    \sqrt{n}[\TP(\hat c_t,\beta^*)- \TP(c^*_t,\beta^*)]&=&f_1^*(c^*_t)\sqrt{n}(\hat c_t-c^*_t)+R_{3n}(t)\nonumber\\
    &=&f_1^*(c^*_t)\sqrt{n}[\widehat{FP}_{\hat\beta}^{-1}(t)- \FP^{-1}_{\beta^*}(t)]+R_{3n}(t),\label{eq: R3}
\end{eqnarray}
%As shown in Lemma \ref{lm: unif R} in Appendix~A, 
The remainder term $R_{3n}(t)$ converges in probability to zero uniformly over the interval 
\[
\big\{t: c^*_t\in[c_{0,L}, c_{0,U}]\big\}=\big[\FP(c_{0,U}, \beta^*), \FP(c_{0,L},\beta^*)\big],
\]
where $c_{0,L}$ and $c_{0,U}$ are specified in Assumption~\ref{assn: cond densities}(iii). 
%As $c_t$ is monotone increasing in $t$, this is an interval of the form $[\delta, 1-\delta]$ for some $\delta\ge 0$.
The asymptotic distribution of the process $t\mapsto \sqrt{n}[\widehat{FP}_{\hat\beta}^{-1}(t)- \FP^{-1}_{\beta^*}(t)]$ can be analyzed in two steps: First, we establish an asymptotically linear representation for the ``base process'' $c\mapsto \sqrt{n}[\widehat{FP}(c,\hat\beta)-\FP(c,\beta^*)]$ that holds uniformly in $c$ (and implies a mean zero Gaussian limit process). Second, we apply the functional delta method under the inverse functional $\phi(F)=F^{-1}$ to characterize the contribution of the term (\ref{eq: R3}) to the asymptotic distribution of the empirical ROC curve. 

Lemma~\ref{lm: FP delta meth} summarizes and completes the development of the approximations presented in equations (\ref{eq: R1}), (\ref{eq: R2}) and (\ref{eq: R3}).  

\begin{lemma}\label{lm: FP delta meth} 
Suppose that Assumptions \ref{assn: iid}, \ref{assn: beta-est}, \ref{assn: empirical process}, \ref{assn: cond densities} and \ref{assn: gradient unif} are satisfied. Then:

(i) $\sup_{t\in [0,1]}R_{1n}(t)=o_p(1)$;

(ii) $\sup_{t\in [0,1]}R_{2n}(t)=o_p(1)$;

(iii) $\sup_{t\in T} R_{3n}(t)=o_p(1)$, where $T=\big[\FP(c_{0,U}, \beta^*), \FP(c_{0,L},\beta^*)\big]$;

(iv) $\widehat{\FP}(c,\hat\beta)$ admits asymptotically linear representation that holds uniformly in $c$:
\begin{equation}\label{eq: R4}
\sqrt{n}\big(\widehat{\FP}(c,\hat\beta)-\FP(c,\beta^*)\big)=\frac{1}{\sqrt{n}}\sum_{i=1}^n \psi_{\FP}(Y_i,X_i,c,\beta^*)+R_{4n}(t),
\end{equation}
where $\sup_{c_\in[a_0,b_0]}R_{4n}(c)=o_p(1)$;
%(ii) The process $c\mapsto \frac{1}{\sqrt{n}}\sum_{i=1}^n \psi_{\FP}(Y_i,X_i,c,\beta^*)$ is stochastically equicontinuous over $[a_0,b_0]$. Hence, $\sqrt{n}[\widehat{\FP}(c,\hat\beta)-\FP(c,\beta^*)]$ converges weakly to a zero mean Gaussian process with covariance kernel $h_{FP}(c_1,c_2)=E[\psi_{\FP}(Y_i,X_i,c_1,\beta^*)\psi_{\FP}(Y_i,X_i,c_2,\beta^*)]$. 

(v) and, by the functional delta method, 
\begin{equation}\label{eq: R5}
\sqrt{n}[\widehat{FP}_{\hat\beta}^{-1}(t)- \FP^{-1}_{\beta^*}(t)]= -\frac{1}{f_0^*(c^*_t)}\sqrt{n}[\widehat{FP}(c^*_t, \hat\beta)- \FP(c^*_t,\beta^*)]+R_{5n}(t),
\end{equation}
where $\sup_{t\in(0,1)}|R_{5n}(t)|=o_p(1)$.
\end{lemma}

\paragraph{Remarks}
\begin{enumerate}
    %\item The weak convergence in part (ii) takes place in $L^\infty([a_0,b_0])$, the space of bounded functions over $[a_0,b_0]$, in the sense of Definition 1.3.3 of van der Vaart and Wellner (1996).
    
    \item The proof of Lemma~\ref{lm: FP delta meth} is provided in Appendix~B; it simply adds some technical details to the arguments outlined in the main text. 
    
    \item The fact that $f^*_0(c)$ is bounded away from zero (Assumption \ref{assn: cond densities}(i)) plays a critical role in ensuring that the remainder term $R_{5n}(t)$ associated with the delta method converges to zero uniformly over the entire $[0,1]$ interval.
    %\item Naturally, an analogous result holds for $\widehat{\TP}(c,\hat\beta)$.
\end{enumerate}

Combining equations (\ref{eq: func decomp}) through (\ref{eq: R5}) with the influence function representations of $\sqrt{n}[\widehat{\TP}(c,\beta^*)-\TP(c,\beta^*)]$ and $\sqrt{n}(\hat\beta-\beta^*)$ yields the following proposition, which is the central result of the paper. 

%\begin{proposition}
%\label{prop: ROC curve asymptotics} Suppose that Assumptions \ref{assn: iid}-\ref{assn: gradient} and \ref{assn: cond densities} are satisfied. Define
%\begin{align}
%%&\psi_{FN}(y,x,c,\beta)=\frac{y}{\pi}\Big( 1\big(G(x,\beta)\leq c\big) -\FN(c,\beta)\Big) + \nabla_\beta \FN(c,\beta)\psi_\beta(y,x,\beta)\label{eq: influ-ROC} ,\\
%%&\psi_{TN}(y,x,c,\beta)=\frac{1-y}{1-\pi}\Big( 1\big(G(x,\beta)\leq c\big) -\TN(c,\beta)\Big) + \nabla_\beta \TN(c,\beta)\psi_\beta(y,x,\beta)\nonumber\\
%&\psi_{R}(y,x,t,\beta^*)=\psi_{TP}(y,x,c^*_t,\beta^*)-
%\frac{f_{1}^*(c^*_t)}{f_{0}^*(c^*_t)}\psi_{FP}(y,x,c^*_t,\beta^*),\nonumber
%\end{align}
%where $c^*_t=\FP^{-1}_{\beta^*}(t)$. Then:
%\begin{align}
%\sup_{t\in T}\Big|\sqrt{n}&\big(\widehat{R}(t,\hat\beta)-R(t,\beta^*)\big)-\frac{1}{\sqrt{n}}\sum_{i=1}^n \psi_{R}(Y_i,X_i,t,\beta^*)\Big|=o_p(1).
%\label{eq: ROC-hat-linear}
%\end{align}
%Furthermore, 
%\begin{align*}
%\sqrt{n}(\widehat{R}_n(t,\hat\beta)-{R}(t,\beta^*))\Rightarrow \Psi_{h_{R}}(t)\text{ in the space $L^\infty(T)$},
%\end{align*}
%where $\Psi_{h_R}(\cdot)$ is a zero mean Gaussian process defined on $T$ with covariance kernel $h_{R}(t_1,t_2)=E[\psi_{R}(Y,X,t_1,\beta^*)\psi_{R}(Y,X,t_2,\beta^*)]$, and ``$\Rightarrow$'' denotes weak convergence. 
%%according to Definition 1.3.3 of van der Vaart and Wellner (1996). 
%%for $t_1,t_2\in[0,1]$.
%\end{proposition}

\begin{proposition}
\label{prop: ROC curve asymptotics} Suppose that Assumptions \ref{assn: iid}, \ref{assn: beta-est}, \ref{assn: empirical process}, \ref{assn: cond densities} and \ref{assn: gradient unif} are satisfied. Define
\begin{align}
&\psi_{R}(y,x,t,\beta^*)=\psi_{TP}(y,x,c^*_t,\beta^*)-
\frac{f_{1}^*(c^*_t)}{f_{0}^*(c^*_t)}\psi_{FP}(y,x,c^*_t,\beta^*),\nonumber
\end{align}
where $c^*_t=\FP^{-1}_{\beta^*}(t)$. Then:

(i) The empirical ROC curve admits an asymptotically linear representation that holds uniformly over $T=\big[\FP(c_{0,U}, \beta^*), \FP(c_{0,L},\beta^*)\big]$:
\begin{align}
\sup_{t\in T}\Big|\sqrt{n}&\big(\widehat{R}(t,\hat\beta)-R(t,\beta^*)\big)-\frac{1}{\sqrt{n}}\sum_{i=1}^n \psi_{R}(Y_i,X_i,t,\beta^*)\Big|=o_p(1),
\label{eq: ROC-hat-linear}
\end{align}
where $c_{0,L}$ and $c_{0,U}$ are chosen in accordance with Assumption~\ref{assn: cond densities}(iii), i.e., $f_1^*/f_0^*$ is continuous and bounded away from zero on $[c_{0,L},c_{0,U}]$ .

(ii) The process $t\mapsto \frac{1}{\sqrt{n}}\sum_{i=1}^n \psi_{R}(Y_i,X_i,t,\beta^*)$ is stochastically equicontinuous over $T$.

(iii) Therefore, 
\begin{align*}
\sqrt{n}(\widehat{R}_n(t,\hat\beta)-{R}(t,\beta^*))\Rightarrow \Psi_{h_{R}}(t)\text{ in the space $L^\infty(T)$},
\end{align*}
where ``$\Rightarrow$'' denotes weak convergence, $L^\infty(T)$ is the space of bounded functions over $T$, and $\Psi_{h_R}(\cdot)$ is a zero mean Gaussian process defined on $T$ with covariance kernel $h_{R}(t_1,t_2)=E[\psi_{R}(Y,X,t_1,\beta^*)\psi_{R}(Y,X,t_2,\beta^*)]$. 

%\begin{itemize}
%    \item [i.] $T=[0,1]$ if Assumption~\ref{assn: cond densities}(ii) holds for $\epsilon=0$; 
%    \item [ii.] $T=[0,\FP(a_1+\epsilon,\beta^*)]$ if Assumption~\ref{assn: cond densities}(ii) holds for $\epsilon>0$ and $b_0<b_1-\epsilon$;
%    \item [iii.] $T=[\FP(b_1-\epsilon,\beta^*),\FP(a_1+\epsilon,\beta^*)]$ if Assumption~\ref{assn: cond densities}(ii) holds for $\epsilon>0$ and $b_1-\epsilon<b_0$.
 %\end{itemize}

\end{proposition}

\paragraph{Remarks}

\begin{enumerate}

\item The precise notion of weak convergence employed in part (iii) is given by Definition 1.3.3 of van der Vaart and Wellner (1996).

\item Given the arguments leading up to Proposition~\ref{prop: ROC curve asymptotics}, the proof of part (i) is practically complete (technically, it still requires showing that the influence function representation of $\sqrt{n}[\widehat{\TP}(c,\beta^*)-\TP(c,\beta^*)]$ holds uniformly in $c$, but this is essentially covered by Lemma~\ref{lm: FP delta meth}(iv)). The proof of Part (ii) relies on Assumptions~\ref{assn: empirical process}, \ref{assn: cond densities} and \ref{assn: gradient unif}. Part (iii) follows immediately from parts (i) and (ii). Details are presented in Appendix~B.

\end{enumerate}

\subsection{Simulating the asymptotic distribution of the ROC curve}

In order to employ Proposition~\ref{prop: ROC curve asymptotics} for statistical inference, we need a method to approximate $\Psi_{h_{R}}(t)$, the distributional limit of the process $\sqrt{n}(\widehat{R}_n(t,\beta^*)-{R}(t,\beta^*))$. To this end, offer two methods: the weighted bootstrap as in Ma and Korosok (2005) and the multiplier bootstrap as in Hsu (2016).

We first present the discussion of the weighted bootstrap. Define $\widehat{\TP}^w(c,\hat{\beta}^w)$ and $\widehat{\FP}^w(c,\hat{\beta}^w)$ precisely as in Section \ref{subsec: weighted} and let $\hat c^w_t= \big(\widehat{\FP}^w_{\hat{\beta}^w}\big)^{-1}(t)$. We can then construct the weighted ROC curve and its estimated limit process as 
$\widehat{R}^w_n(t,\hat\beta^w)=\widehat{\TP}^w(\hat{c}^w_t,\hat{\beta}^w)$ and $\widehat{\Psi}^w_{R,n}(t)=\sqrt{n}(\widehat{R}^w_n(t,\hat\beta^w)- \widehat{R}_n(t,\hat\beta))$.

\begin{proposition}\label{prop: weighted bootstrap}
Suppose that \ref{assn: iid}-\ref{assn: gradient}, and
\ref{assn: weighted bootstrap W}-\ref{assn: cond densities} are satisfied. Then, conditional on the sample path of the data,
$\widehat{\Psi}^w_{R,n}(\cdot)\Rightarrow \Psi_{h_{R,2}}(\cdot)$  in the space $L^\infty(T)$ with probability approaching one.
% (w.p.a.\ 1) which is denoted by $\widehat{\Psi}^u_{R,n}(t)\stackrel{p}{\Rightarrow}\Psi_{h_{R,2}}(t)$. 
%\label{thm: simulated-process}
\end{proposition}

Under the conditions of Proposition \ref{prop: ROC curve asymptotics}, one can apply the arguments in Theorem 2 of Ma and Kosorok (2005) to show that $\widehat{\Psi}^w_{R,n}(t)$ also approximates the distribution of $\Psi_{h_{R,2}}(t)$ in the sense of Proposition \ref{prop: multiplier bootstrap}. That is, conditional on the sample path of the data,
$\widehat{\Psi}^w_{R,n}(t)\Rightarrow \Psi_{h_{R,2}}(t)$  with probability approaching one.

We now turn to the discussion of the multiplier bootstrap method that is based on the conditional multiplier central limit theorem (see, e.g., van der Vaart and Wellner 1996, Section~2.9). The method requires consistent estimation of the components of the influence function $\psi_R$, uniformly in $t$. However, this estimation needs to be performed only once, over the original data set, given that the method does not rely on successive resampling and reestimation. 

Let $\widehat{\psi}_\beta(y,x, \hat{\beta})$ denote the estimated influence function of $\hat\beta$, where we replace any unknown parameters or functions within $\psi_\beta$ with consistent estimators (note that this function does not depend on $t$). We make the general assumption that the asymptotic variance-covariance matrix of $\hat\beta$ is consistently estimable using $\widehat{\psi}_\beta$ and the sample analog principle:

\begin{assumption}\label{assn: consistent V}
Let $V=E[{\psi}_\beta(Y,X, \beta^*){\psi}_\beta(Y,X, \beta^*)']$. Then: 
\begin{align*}
\widehat{V}_n=n^{-1}\sum_{i=1}^n
\widehat{\psi}_\beta(Y_i,X_i, \hat{\beta}_n)\widehat{\psi}_\beta(Y_i,X_i,
\hat{\beta}_n)'\stackrel{p}{\rightarrow} V.
\end{align*}
%where $V=E[{\psi}_\beta(Y,X, \beta^*){\psi}'_\beta(Y,X, \beta^*)]$.
\end{assumption}
%Assumption \ref{assu: consistent V} requires that we have a
%consistent estimator for the asymptotic covariance matrix of the estimator $\hat{\beta}_n$.

%Let $c_t$ denote the $t$-th quantile of  the conditional distribution of  $G(X,\beta^*)$ given $Y=0$. 
Let $\mathcal{C}=[c_{0,L},c_{0,U}]$.  We further assume that there exist uniformly consistent estimators for $\nabla_\beta \FP(c,\beta^*)$, $\nabla_\beta \TP(c,\beta^*)$ and $f_{1}^*(c)/f_{0}^*(c)$ on $\mathcal{C}$. Here we state the existence of these estimators as a high level assumption and provide concrete implementations and additional assumptions in Appendix C.  
%\begin{assumption}\label{assn: consistent estimator}
%The estimators $\nabla_\beta
%\widehat{\FN}(c,\hat{\beta}_n)$,  $\nabla_\beta
%\widehat{\TN}(c,\hat{\beta}_n)$, $\hat{f}_{1}(c)$ and
%$\hat{f}_{0}(c)$ satisfy 
%\begin{align*}
%&\sup_{c\in\mathcal{C}}\|\nabla_\beta\widehat{\FN}(c,\hat{\beta}_n)-\nabla_\beta \FN(c,\beta^*)\|=o_p(1),\\
%&\sup_{c\in\mathcal{C}}\|\nabla_\beta\widehat{\TN}(c,\hat{\beta}_n)-\nabla_\beta \TN(c,\beta^*)\|=o_p(1),\\
%&\sup_{c\in\mathcal{C}}|\hat{f}_{1}(c)-f_{1}^*(c)|=o_p(1), ~~~\sup_{c\in\mathcal{C}}|\hat{f}_{0}(c)-f_{0}^*(c)|=o_p(1).
%\end{align*}
%\end{assumption}

\begin{assumption}\label{assn: consistent estimator}
The estimators $\nabla_\beta
\widehat{\FP}(c,\hat{\beta}_n)$,  $\nabla_\beta
\widehat{\TP}(c,\hat{\beta}_n)$, and $\hat{f}_{1}(c)/\hat{f}_{0}(c)$  are Lipschitz continuous in $c$ on  
$\mathcal{C}$ which is compact and satisfy 
\begin{align*}
&\sup_{c\in\mathcal{C}}\|\nabla_\beta\widehat{\FP}(c,\hat{\beta}_n)-\nabla_\beta \FP(c,\beta^*)\|=o_p(1),\\
&\sup_{c\in\mathcal{C}}\|\nabla_\beta\widehat{\TP}(c,\hat{\beta}_n)-\nabla_\beta \TP(c,\beta^*)\|=o_p(1),\\
&\sup_{c\in\mathcal{C}}\left|\frac{\hat{f}_{1}(c)}{\hat{f}_{0}(c)}-\frac{f_{1}(c)}{f_{0}(c)}\right|=o_p(1).
\end{align*}
In addition, the estimator $\hat{c}_t$ is uniformly consistent for $c_t$ for $t\in T$.
\end{assumption}

We now present the multiplier bootstrap. Let $U_1,\ldots, U_n$ be i.i.d.\ random variables independent of the data with moments $E[U]=0$, $E[U^2]=1$, and $E|U|^{2+\delta_u}<\infty$ for some $\delta_u>0$. For $t\in[0,1]$, we define the simulated stochastic process $\widehat{\Psi}^u_{R,n}(t)$ as
\begin{align}
\widehat{\Psi}^u_{R,n}(t)=\frac{1}{\sqrt{n}}\sum^n_{i=1}U_i\cdot\widehat{\psi}_R(Y_i,X_i,t,\hat\beta),
\label{eq: simu-psi}
\end{align}
where
\begin{align*}
&\hat{\psi}_{R}(y,x,t,\beta)=\hat{\psi}_{TP}(y,x,\hat{c}_{t},\beta)
-\frac{\hat{f}_{1}(\hat c_t)}{\hat{f}_{0}(\hat{c}_{t})}\hat{\psi}_{FP}(y,x,\hat{c}_{t},\beta),\\
&\hat{\psi}_{TP}(y,x,c,\beta)=\frac{y}{\hat{\pi}}\Big(1[G(x,\beta)> c] - \widehat{\TP}(c,\beta)\Big)+
\nabla_\beta \widehat{\TP}(c,\beta)\hat{\psi}_\beta(y,x,\beta) ,\\
&\hat{\psi}_{FP}(y,x,c,\beta)=\frac{1-y}{1-\hat{\pi}}\Big(1[G(x,\beta)> c] - \widehat{\FP}(c,\beta)\Big)+
\nabla_\beta \widehat{\FP}(c,\beta)\hat{\psi}_\beta(y,x,\beta) ,\\
&\hat{\pi}=\frac{1}{n}\sum_{i=1}^n Y_i.
\end{align*}

The next result shows that the distribution of the simulated process $\widehat{\Psi}^u_{R,n}(t)$ approximates that of the true limiting process $\Psi_{h_{R,2}}(t)$ in large samples.

\begin{proposition}\label{prop: multiplier bootstrap}
Suppose that Assumptions \ref{assn: iid}-\ref{assn: gradient} and \ref{assn: cond densities}-\ref{assn: consistent estimator} are satisfied. Then, conditional on the sample path of the data,
$\widehat{\Psi}^u_{R,n}(\cdot)\Rightarrow \Psi_{h_{R,2}}(\cdot)$  in the space $L^\infty(T)$ with probability approaching one.
% (w.p.a.\ 1) which is denoted by $\widehat{\Psi}^u_{R,n}(t)\stackrel{p}{\Rightarrow}\Psi_{h_{R,2}}(t)$. 
\label{thm: simulated-process}
\end{proposition}

The weighted bootstrap has the advantage that it does not require explicit estimation of this function, but it is computationally somewhat more costly. On the other hand, the proof of weighted bootstrap is less involved because the multiplier method relies heavily on Assumption \ref{assn: consistent estimator}, i.e., the availability of uniformly consistent estimators for the components of $\psi_R$. To obtain estimators satisfy  Assumption \ref{assn: consistent estimator} additional assumptions are needed and in Appendix C, we provide estimators and additional assumptions so that Assumption \ref{assn: consistent estimator} can be satisfied and we can apply multiplier method.

\section{Applications to various inference problems}\label{sec: inference}

In this section, we provide some examples that we can apply the results in Section \ref{sec: uniform results}.

\subsection{Uniform confidence bands}\label{subsec: uniform CB}
Let $\widehat{\sigma}^2_t$ 
%$\widehat{\sigma}^2_t=n^{-1}\sum_{i=1}^n \widehat{\psi}^2_R(Y_i,X_i,t,\hat\beta)$ 
denote a uniform consistent estimator for ${\sigma}^2_t$, the asymptotic variance of $\sqrt{n}(\widehat{R}_n(t,\hat\beta)-{R}(t,\beta^*))$ for $t\in T$.  Later we will provide two estimators based on weighted bootstrap method and analytic results. 
Let $\widehat{\sigma}_{t,\epsilon}=\max\{\widehat{\sigma}_t,\epsilon\}$ in which $\epsilon>0$ is a fixed and small number. We are interested in a standardized version of confidence bands and by truncating $\widehat{\sigma}_t$ by $\epsilon$, we can make sure that we will not divide something close to zero when $t$ is close to 0 or 1.  

For a nominal significance level $\alpha$ and for $\tau_\ell,\tau_u\in T$ with $\tau_\ell\leq \tau_u$, let $\widehat{C}^\text{1-sided}_\alpha$ and $\widehat{C}^\text{2-sided}_\alpha$ respectively denote the one- and two-sided critical values that satisfy
\begin{align}
\widehat{C}^\text{1-sided}_\alpha&=\inf_{ a \in R}\left\{P\left(\sup_{t\in[\tau_\ell,\tau_u]}
\frac{\widehat{\Psi}^w_{R,n}(t) }{\widehat{\sigma}_{t,\epsilon}}\leq a\right)\geq1-\alpha\right\},
\label{eq: one side CV}\\
\widehat{C}^\text{2-sided}_\alpha&=\inf_{a \in R}\left\{P\left(\sup_{t\in[\tau_\ell,\tau_u]}\frac{|\widehat{\Psi}^w_{R,n}(t)|}{\widehat{\sigma}_{t,\epsilon}}\leq a\right)\geq1-\alpha\right\}.\label{eq: two side CV}
\end{align}
Here, $\widehat{C}^\text{1-sided}_\alpha$ and $\widehat{C}^\text{2-sided}_\alpha$ are, respectively, the $(1-\alpha)$th quantile of $\sup_{t\in[\tau_\ell,\tau_u]}\widehat{\Psi}^w_{R,n}(t) \big/\widehat{\sigma}_{t,\epsilon}$ and $(1-\alpha)$th quantile of $\sup_{t\in[\tau_\ell,\tau_u]}\big|\widehat{\Psi}^w_{R,n}(t) \big/\widehat{\sigma}_{t,\epsilon}\big|$.  Note that one can replace  $\widehat{\Psi}^w_{R,n}(t) $ with $\widehat{\Psi}^u_{R,n}(t) $ to 
construct $\widehat{C}^\text{1-sided}_\alpha$ and $\widehat{C}^\text{2-sided}_\alpha$ as well. 

Once the critical values are constructed, we can also obtain one- and two-sided uniform confidence bands for ${R}(t,\beta^*)$ over $[\tau_\ell,\tau_u]$. Specifically, the one-sided $(1-\alpha)$ uniform confidence band is given by
\begin{equation}
\label{band:one}
\left(\widehat{R}_n(t,\hat\beta)-\widehat{C}^\text{1-sided}_\alpha\frac{\widehat{\sigma}_{t,\epsilon}}{\sqrt{n}},\quad+\infty\right),\quad\tau\in[\tau_\ell,\tau_u]
\end{equation}
and the two-sided $(1-\alpha)$ uniform confidence band is
\begin{equation}
\label{band:two}
\left(\widehat{R}_n(t,\hat\beta)-\widehat{C}^\text{2-sided}_\alpha\frac{\widehat{\sigma}_{t,\epsilon}}{\sqrt{n}},\quad\widehat{R}_n(t,\hat\beta)+\widehat{C}^\text{2-sided}_\alpha\frac{\widehat{\sigma}_{t,\epsilon}}{\sqrt{n}}\right),\quad\tau\in[\tau_\ell,\tau_u].
\end{equation}

\bigskip
\noindent {\bf Implementation of Uniform Confidence Bands}

We now provide a step-by-step implementation for constructing uniform confidence bands.
\begin{enumerate}
\item
Obtain $\widehat{R}_n(t,\hat\beta)$ from Section~\ref{sec: uniform results} and $\widehat{\sigma}_{t,\epsilon}$ from Section \ref{subsec: uniform CB} with $t\in\{\tau_\ell,\tau_\ell+0.01,\dotsc,\tau_u\}$.
\item
Draw i.i.d.\ pseudo random variables $\{W_1,\dotsc,W_n\}$ where $W_i$'s are normal distributions with mean and variance equal to one $B$ times for, say, $B=1000$. For each repetition $b=1,\dotsc,B$, calculate the simulated process $\widehat{\Psi}^w_{R,n}(t)$ according to (\ref{eq: simu-psi}).
\item
For the one-sided case, store the maximum value of ${\widehat{\Psi}^w_{R,n}(t) }\big/{\widehat{\sigma}_{t,\epsilon}}$ over the grid of $t$ values set up in Step 1; that is, let $M_b=\max_{t\in\{\tau_\ell,\tau_\ell+0.01,\dotsc,\tau_u\}}{\widehat{\Psi}^w_{R,n}(t) }\big/{\widehat{\sigma}_{t,\epsilon}}$ for $b=1,\dotsc,B$.
\item
Rank the $M_b$ values in an ascending order so that $M_{(1)}\leq\dotsc\leq M_{(B)}$. Next, define $M_{(\lfloor(1-\alpha)B\rfloor)}$ as the critical value $\widehat{C}^\text{1-sided}_\alpha$, where $\lfloor a \rfloor$ is the floor function returning the largest integer not greater than $a$. The one-sided $(1-\alpha)$ uniform confidence bands for $\{{R}(t,\beta^*):t\in[\tau_\ell,\tau_u]\}$ are given by \eqref{band:one}.
\item
For the two-sided case, simply replace ${\widehat{\Psi}^w_{R,n}(t) }\big/{\widehat{\sigma}_{t,\epsilon}}$ in Step 3 with $\big|{\widehat{\Psi}^w_{R,n}(t) }\big|\big/{\widehat{\sigma}_{t,\epsilon}}$ and repeat Step 4 for the critical value $\widehat{C}^\text{2-sided}_\alpha$. The two-sided $(1-\alpha)$ uniform confidence band for $\{{R}(t,\beta^*):t\in[\tau_\ell,\tau_u]\}$ is given by \eqref{band:two}.
\end{enumerate}

\bigskip
\noindent {\bf  Uniformly consistent estimators for ${\sigma}^2_t$}\\
We consider two estimators here.  First estimator is based on weighted bootstrap that is similar to Remark 2 after Proposition \ref{prop: est effect gen-bootstrap}. 
Let $\widehat{\Psi}^w_{R,n}(t)$ denote the ROC estimate from the $w$th bootstrap cycle, $w=1,\ldots, \mathcal{W}$. Then one can estimate ${\sigma}^2_t$ by    
\begin{align*}
 &\widehat{\sigma}^2_t=\frac{n}{\mathcal{W}}\sum_{w=1}^\mathcal{W} \big(\widehat{R}^w_n(t,\hat\beta^w)-\overline{\widehat{R}}^w_n(t,\hat\beta^w)\big)\big(\widehat{R}^w_n(t,\hat\beta^w)-\overline{\widehat{R}}^w_n(t,\hat\beta^w)\big)',\text{ where}\\
   &\overline{\widehat{R}}^w_n(t,\hat\beta^w)=\frac{1}{\mathcal{W}}\sum_{w=1}^\mathcal{W}\widehat{R}^w_n(t,\hat\beta^w).
\end{align*}
We have that conditional on sample path with probability approaching one,
\begin{align*}
&\widehat{\sigma}^2_t \stackrel{p}{\rightarrow}_w
\frac{1}{n}\sum_{i=1}^n \psi^2_{R}(Y_i,X_i,t,\beta^*)+o_p(1),
\end{align*}
where $\stackrel{p}{\rightarrow}_w$ denotes probability limit under the law of the $W_i$'s. It follows that uniformly over $t\in[t_\ell,t_u]$,
\begin{align*}
\lim_{\mathcal{W} \rightarrow \infty} \widehat{\sigma}^2_t\stackrel{p}{\rightarrow} {\sigma}^2_t.
\end{align*}

The second estimator is based on analytic results. Recall that $\widehat{\psi}_R(Y_i,X_i,t,\hat\beta)$ is the estimated influence function for $\widehat{R}_n(t,\hat\beta)$ used in the multiplier bootstrap method. A uniformly consistent estimator for $ {\sigma}^2_t$ is given by
\begin{align*}
\widehat{\sigma}^2_t=\frac{1}{n}\sum_{i=1}^n \widehat{\psi}^2_R(Y_i,X_i,t,\hat\beta)
\end{align*}
and this is shown in the proof of \ref{prop: multiplier bootstrap}.  

\subsection{ROC dominance test}
\label{subsec: ROC dominance}
For two predictive index models $G_1(X,\beta_1)$ and $G_2(X,\beta_2)$, we may want to test whether $G_1$ has strictly better predictive power than $G_2$ in the sense that the ROC curve associated with $G_1$ dominates the ROC curve associated with $G_2$. What domination means is that for any given false positive rate $G_1$ delivers a higher true positive rate, i.e., the ROC curve for $G_1$ always lies above the ROC curve for $G_2$. Any decision maker, regardless of their loss function and their optimal cutoff, would then prefer model $G_1$ over $G_2$. 

Let $R_1(t,\beta_1^*)$ and $R_2(t,\beta_2^*)$ denote the ROC curves associated with $G_1$ and $G_2$, respectively. The hypotheses that $R_1(t,\beta_1^*)$ dominates $R_2(t,\beta_2^*)$ can be formally stated as
\begin{align}
&H_0:  R_2(t,\beta_2^*)\leq R_1(t,\beta_1^*) ~~\text{ for all $t\in[0,1]$},\nonumber\\
&H_1:  R_2(t,\beta_2^*) >R_1(t,\beta_1^*) ~~\text{ for some $t\in[0,1]$}. \label{eq: null-ROC-Dominace-1}
\end{align}
Our test for ROC dominance is similar to the test for first order stochastic dominance in Barrett and Donald (2003) and Donald and Hsu (2016) except that we need to consider the estimation effect of $\hat{\beta}$ as in Linton, Massoumi and Whang (2005) and Linton, Song and Whang (2010). 

Let $\widehat{R}_{j,n}(t,\hat{\beta}_j)$  be the estimators
for $R_j(t,\beta_j^*)$ for $j=1,2$.  
Define $\psi_{j,R}(Y_i,X_i,t,\beta_1^*)$ for $j=1$ and 2 as above.
Let $\widehat{\sigma}^2_{RD}(t)$
%$\widehat{\sigma}^2_{RD}(t)=n^{-1}\sum_{i=1}^n (\widehat{\psi}_{2,R}(Y_i,X_i,t,\hat{\beta}_2)-\widehat{\psi}_{1,R}(Y_i,X_i,t,\hat{\beta}_2))^2$
denote a uniform consistent estimator for ${\sigma}^2_{RD}(t)$, the asymptotic variance of $\sqrt{n}(\widehat{R}_{2,n}(t,\hat{\beta}_2)-\widehat{R}_{1,n}(t,\hat{\beta}_1)-{R}_{2}(t,{\beta}^*_2)-{R}_{1}(t,{\beta}^*_1))$.
Let $\widehat{\sigma}_{RD,\epsilon}(t)=\max\{\widehat{\sigma}_{RD}(t),\epsilon\}$ in which $\epsilon>0$ is a fixed and small number.  Uniform consistent estimator $\widehat{\sigma}^2_{RD}(t)$ can be obtained similar to $\widehat{\sigma}^2_t$ in Section \ref{subsec: uniform CB}, so we omit the details. 

We define the test statistic as
$\widehat{S}_n=\sqrt{n}\sup_{t\in[0,1]}(\widehat{R}_{2,n}(t,\hat{\beta}_2)-\widehat{R}_{1,n}(t,\hat{\beta}_1))/\widehat{\sigma}_{RD,\epsilon}(t)$.
Define the weighted bootstrap process $\Psi^w_{RD,n}(t)$ as 
$\widehat{\Psi}^w_{R,n}(t)=\sqrt{n}\big(\widehat{R}^w_{2,n}(t,\hat\beta^w_2)-\widehat{R}^w_{1,n}(t,\hat\beta^w_1)
-(\widehat{R}_{2,n}(t,\hat{\beta}_2)-\widehat{R}_{1,n}(t,\hat{\beta}_1))\big)
$ and define the multiplier bootstrap process  $\Psi^u_{RD,n}(t)$ as
\begin{align*}
\widehat{\Psi}^u_{RD,n}(t)=\frac{1}{\sqrt{n}}\sum_{i=1}^nU_i\cdot(\widehat{\psi}_{2,R}(Y_i,X_i,t,\hat{\beta}_2)-\widehat{\psi}_{1,R}(Y_i,X_i,t,\hat{\beta}_1)).
\end{align*}

Under the least favorable configuration, we define the weighted bootstrap critical
value as
\begin{align}
\hat{c}_n=\sup \Big\{c\big|P^{w}\Big(\sqrt{n}\sup_{t\in[0,1]}\frac{\widehat{\Psi}^w_{RD,n}(t)}{\widehat{\sigma}_{RD,\epsilon}(t)}\leq c\Big)
\leq 1-\alpha \Big\},
\end{align}
with significance level $\alpha$.  The decision rule is
\begin{align}
\text{Reject $H_0$ if $\widehat{S}_n>\hat{c}_n$.}\label{eq: decision rule}
\end{align}
Then one can use $\widehat{\Psi}^u_{RD,n}(t)$ to construct critical value  $\hat{c}_n$ as well.

Similar to the stochastic dominance test literature, we can show
that under the null hypothesis the asymptotic size of a test with
decision rule defined in (\ref{eq: decision rule}) is less than or
equal to $\alpha$.  That is, we can control the asymptotic size of
our ROC dominance test well.  Also, under the fixed alternative, we have the test statistic converging to positive infinity and the
critical value converging to a finite number, so the test is
consistent.  Our test is based on least favorable
configuration, so it is conservative in that the asymptotic size is strictly smaller than $\alpha$ unless $R_2(t,\beta_2^*)= R_1(t,\beta_1^*)$ for all $t\in[0,1]$.  One can improve the power of our test by using the recentering method in Hansen (2005), Donald and Hsu (201) which is similar to the generalized moment selection method in Andrews and Soreas (2010), and Andrews and Shi (2013), and the contact set approach in Linton, Song and Whang (2010).  In this paper, we do not adopt this approach but the extension is
straightforward.

\subsection{Comparing AUCs}

Recall that AUC is defined as the integral of ROC curve from 0 to one.  
Following Section \ref{subsec: ROC dominance}, let $R_1(t,\beta_1^*)$ and $R_2(t,\beta_2^*)$ denote the ROC curves for two predictive index models $G_1(X,\beta_1)$ and $G_2(X,\beta_2)$. 
Let $AUC_j=\int_{0}^1 R_j(t,\beta_j^*) dt$  and its estimator be $\widehat{AUC}_j=\int_{0}^1 
\widehat{R}_{j,n}(t,\hat{\beta}_j)dt $ for $j=1$ and 2.  Then it is true that
\begin{align*}
\sqrt{n}(\widehat{AUC}_2-\widehat{AUC}_1)= 
\frac{1}{\sqrt{n}}\sum_{i=1}^n \int_{0}^{1}({\psi}_{2,R}(Y_i,X_i,t,\beta^*_2)-{\psi}_{1,R}(Y_i,X_i,t,{\beta}^*_1)dt+o_p(1)
\end{align*} 
and
\begin{align*}
&\sqrt{n}(\widehat{AUC}_2-\widehat{AUC}_1){\rightarrow}_d
N[0, \mathcal{V}_{a}],\\
& \mathcal{V}_{a}=E[\big(\int_{0}^{1} {\psi}_{2,R}(Y,X,t,\beta^*_2)-{\psi}_{1,R}(Y,X,t,{\beta}^*_1)dt\big)^2].
\end{align*}
To make inference, one can use a weighted bootstrap method to approximate the limiting distribution of $N[0, \mathcal{V}_{a}]$ or one can estimate $ \mathcal{V}_{a}$ analytically by 
\begin{align*}
\widehat{\mathcal{V}}_{a}=
\frac{1}{n}\sum_{i=1}^n
\Big(\int_{0}^1(\widehat{\psi}_{2,R}(Y_i,X_i,t,\hat{\beta}_2)-\widehat{\psi}_{1,R}(Y_i,X_i,t,\hat{\beta}_1))dt\Big)^2.
\end{align*}
For brevity, we omit the details here. 

\bigskip
\noindent {\bf Remark}\\
Suppose that $G(X,\beta)$ is a correct specification for the propensity score function, $P(Y=1|X)$, in that there exists $\beta^*$ such that  $G(X,\beta^*)=P(Y=1|X)$ a.s. Then the estimation effect of $\beta^*$ on the distribution of AUC is negligible. It is then true that when two predictive predictive index models, $G_1(X,\beta_1)$ and $G_2(X,\beta_2)$,  are both correctly specified for $P(Y=1|X)$, we have  $\mathcal{V}_{a}=0$, i.e., the limiting distribution of $\sqrt{n}(\widehat{AUC}_2-\widehat{AUC}_1)$ is degenerate.

\section{Simulations: the relevance of the estimation effect}\label{sec: simulations}

%\subsection{Monte Carlo illustration of the estimation effect}

We now present a small Monte Carlo simulation to illustrate the theoretical discussion of the estimation effect and the pointwise asymptotic results in Section~\ref{sec: pointwise results}. The data generating process (DGP) is specified as follows. Let $X=(X_1, X_2, X_3)'$ be a $3\times 1$ vector of predictors and $\tilde X=(1,X')'$. The components of $X$ are either independent $N(0,1)$ or unif$[-.5,1.5]$ variables. The outcomes $Y$ are generated according to the conditional probability function
\[
p(X)=G(X,\beta^\circ)=G(\tilde X'\beta^\circ)\text{ with } \beta^\circ=(0, 0.5, 0.25, 1)'\text{ and }G\in\{\text{logit},\text{cauchit}\}.
\]
In the majority of the exercises we use a logistic link in the DGP (so that the logit first stage is correctly specified), but we also conduct some simulations with a cauchit link (so that the logit first stage is mildly misspecified). 

Given a sample of observations and a cutoff $c$, we construct nominally 90\% confidence intervals for \TP$(c,\beta^\circ)$ and \TP$(c,\beta^\circ)-\FP(c,\beta^\circ)$ in three different ways: (i) using the true predictive index $p(X)=G(\tilde X'\beta^\circ)$ with the conventional limit distributions (\ref{eq: \TP asy dist}) and (\ref{eq: \FP asy dist}); (ii) using the estimated predictive index $\Lambda(\tilde X'\hat\beta)$ with the conventional limit distributions (so that the estimation effect is ignored); (iii) using the estimated predictive index $\Lambda(\tilde X'\hat\beta)$ with the corrected limit distribution (\ref{eq: TF joint dist}). We simulate 10,000 samples and compute the actual coverage probability of these intervals. 

\begin{table}[thbp]

{\footnotesize

\begin{center}
\topcaption{Illustration of the estimation effect: actual coverage probabilities}

\begin{tabular}{l cccc|cccc}
     &\multicolumn{4}{c|}{Nominal\ 90\% CIs for \TP}&\multicolumn{4}{c}{Nominal\ 90\% CIs for \TP-\FP}\\
     &True&\multicolumn{2}{c}{\underline{Conventional}}&\multicolumn{1}{c|}{\underline{Corrected}}&True&\multicolumn{2}{c}{\underline{Conventional}}&\multicolumn{1}{c}{\underline{Corrected}}\\[-5pt]
     $c$ & value & $G(\tilde X' \beta^\circ)$ & $\Lambda(\tilde X' \hat\beta)$ & $\Lambda(\tilde X' \hat\beta)$  & value\ & $G(\tilde X' \beta^\circ)$ & $\Lambda(\tilde X' \hat\beta)$ & $\Lambda(\tilde X' \hat\beta)$\\ 
     
     \hline
     &\multicolumn{8}{c}{(A) $n=200$, $X_1, X_2, X_3\sim$iid $N(0,1)$}\\
     \hline
0.2  &0.970 &0.794  & 0.793  & 0.885 & 0.166 & 0.897 & 0.628 & 0.850\\[-5pt]
0.33 &0.884 &0.887  & 0.856  & 0.892 & 0.313 & 0.893 & 0.778 & 0.885\\[-5pt]
0.5  &0.694 &0.894  & 0.768  & 0.890 & 0.388 & 0.896 & 0.889 & 
0.896\\[-5pt]
0.67 &0.429 &0.894  & 0.620  & 0.876 & 0.313 & 0.897 & 0.769 & 0.878\\[-5pt]
0.8  &0.196 &0.895  & 0.533  & 0.847 & 0.166 & 0.894 & 0.615 & 0.842\\
\hline
 &\multicolumn{8}{c}{(B) $n=500$, $X_1, X_2, X_3\sim$iid $N(0,1)$}\\
\hline

0.2  & 0.970 & 0.861 & 0.843 & 0.893 & 0.166 &0.905 & 0.625 & 0.862\\[-5pt]
0.33 & 0.884 & 0.895 & 0.861 & 0.891 & 0.313 &0.903 & 0.779 & 0.888\\[-5pt]
0.5  & 0.694 & 0.892 & 0.766 & 0.891 & 0.388 &0.901 & 0.896 & 0.899\\[-5pt]
0.67 & 0.429 & 0.897 & 0.618 & 0.886 & 0.313 &0.900 & 0.773 & 0.886\\[-5pt]
0.8  & 0.196 & 0.895 & 0.531 & 0.862 & 0.166 &0.895 & 0.627 & 0.864\\

\hline
 &\multicolumn{8}{c}{(C) $n=2500$, $X_1, X_2, X_3\sim$iid $N(0,1)$}\\
\hline

0.2  & 0.970 & 0.893 & 0.857 & 0.901 & 0.166 & 0.902 & 0.630 & 0.885\\[-5pt]
0.33 & 0.884 & 0.898 & 0.863 & 0.899 & 0.313 & 0.899 & 0.779 & 0.899\\[-5pt]
0.5  & 0.694 & 0.899 & 0.779 & 0.902 & 0.388 & 0.901 & 0.902 & 0.899\\[-5pt]
0.67 & 0.429 & 0.898 & 0.633 & 0.896 & 0.313 & 0.894 & 0.776 & 0.896\\[-5pt]
0.8  & 0.196 & 0.895 & 0.545 & 0.881 & 0.166 & 0.898 & 0.630 & 0.885\\

\hline
 &\multicolumn{8}{c}{(D) $n=500$, $X_1, X_2, X_3\sim$unif $[-0.5,1.5]$}\\
\hline
0.5 & 0.934 & 0.889 & 0.515 & 0.876 &0.117 & 0.888 & 0.628 &0.855\\[-5pt]
0.67& 0.671 & 0.897 & 0.595 & 0.904 &0.257 & 0.894 & 0.897 &0.926\\[-5pt]
0.8 & 0.304 & 0.899 & 0.378 & 0.862 &0.182 & 0.894 & 0.675 &0.899\\

\hline
 &\multicolumn{8}{c}{(E) $n=500$, $X_1, X_2, X_3\sim$iid $N(0,1)$, $G$=cauchit}\\
\hline
0.2  & 0.963 & 0.878 & 0.841 & 0.879 & 0.157 &0.899 &0.602 & 0.862\\[-5pt]
0.33 & 0.862 & 0.902 & 0.626 & 0.657 & 0.339 &0.898 &0.701 & 0.858\\[-5pt]
0.5  & 0.702 & 0.898 & 0.769 & 0.898 & 0.404 &0.896 &0.890 & 0.895\\[-5pt]
0.67 & 0.476 & 0.901 & 0.494 & 0.814 & 0.339 &0.903 &0.701 & 0.857\\[-5pt]
0.8  & 0.194 & 0.895 & 0.523 & 0.861 & 0.157 &0.899 &0.601 & 0.858\\
\hline

%Conventional 90% CI with estimated beta: cov. prob. for \TP pseudo true value

%For c=0.2 cov. prob.=0.8683
%For c=0.33 cov. prob.=0.8827
%For c=0.5 cov. prob.=0.7699
%For c=0.67 cov. prob.=0.5914
%For c=0.8 cov. prob.=0.5227

%Conventional 90% CI with estimated beta: cov. prob. for \TP-\FP pseudo true value
 
%For c=0.2 cov. prob.=0.5992
%For c=0.33 cov. prob.=0.7432
%For c=0.5 cov. prob.=0.8899
%For c=0.67 cov. prob.=0.7415
%For c=0.8 cov. prob.=0.6039

\end{tabular}

\label{tbl: est effect}
\end{center}
Note: $c$ is the cutoff; ``True value'' is the true value of $\TP(c,\beta^\circ)$ and $\TP(c,\beta^\circ)$-$\FP(c,\beta^\circ)$. All other numbers in the table are actual coverage probabilities. $G(\tilde X'\beta^\circ)$ means using the true value of $\P(Y=1|X)$ as the predictive index; $\Lambda(\tilde X'\hat\beta)$ means pre-estimating the predictive index by a logit regression of $Y$ on $\tilde X=(1,X')'$. The columns labeled ``Conventional'' report CIs based on the limit distributions (\ref{eq: \TP asy dist}) and (\ref{eq: \FP asy dist}). The columns labeled ``Corrected'' report CIs based on (\ref{eq: TF joint dist}), which accounts for the pre-estimation effect.  

}
\end{table}

Table \ref{tbl: est effect} reports the results from this exercise for $c\in\{0.2, 0.33, 0.5, 0.67, 0.8\}$ and $n\in\{200, 500, 5000\}$. The first message is that failing to account for the pre-estimation effect can cause substantial distortions in the coverage probability of the conventional CIs. In panels (A) through (D) the estimation effect can be seen by comparing the columns titled ``Conventional $G(\tilde X'\beta^\circ)$'' and  ``Conventional $\Lambda(\tilde X'\hat\beta)$.'' In the former case there is no estimation effect and any deviation from the nominal confidence level of 90\% is a small sample phenomenon.\footnote{For example, for $c=0.2$ the value of \TP$(c,\beta^\circ)$ is close to the upper bound 1, and the coverage probability of the fixed-$\beta$ CI is only 80\% for $n=200$.} Over the various cases, the estimation effect ranges from  essentially zero to as large as a 30 to 40 percentage point difference in coverage probability. In panel (E), the comparison between the same two columns includes the estimation effect as well as some ``bias'' due to the fact that the first stage logit regression is misspecified. 

The theory presented in Section~\ref{subsec: est eff simple theory} gives insight into why the estimation effect is negligible in some cases. In particular, consider the parameter $TP-FP$ on panels (A) through (C) with $c=0.5$. As the predictors are independent standard normal variables and there is no constant in the DGP, the symmetry of the logistic cdf gives $\pi=\P(Y=1)=0.5$. Therefore, when $c=0.5$, $TP-FP$ is a scalar multiple of $(1-c)\pi\TP-c(1-\pi)\FP$. As explained in footnote~\ref{fn: no est effect}, inference about this particular linear combination is not impacted by the pre-estimation effect. This is clearly reflected in the estimation results. By contrast, in panel (D) the predictor distribution is not symmetric around zero, so $\pi\neq 0.5$, and the estimation effect is indeed present for $TP-FP$ even when $c=0.5$.       

The second main message is that the proposed analytical correction works well in virtually all the cases considered here. This includes panel (A), where the sample size is small, and panel (E), where the first stage logit model is misspecified. Not surprisingly, under misspecification the corrected CI can also fall somewhat short of the 90\% confidence level, but it still represents a large improvement over conventional inference.   

\section{Conclusions}\label{sec: concl}

We provided both pointwise and uniform asymptotic results that describe the distribution of an empirical ROC curve based on a pre-estimated index. The core theory is complete. Ongoing work consists of: (i) developing appropriate test procedures when the first stage models are nested and the ROC influence functions are the same under the null; (ii) additional simulations that illustrate the small sample performance of the uniform asymptotic results, the practical use of the tests, and the power gains afforded by in-sample inference.

\newpage

\begin{apabib}
Abadie, A.\ and  G.W.\ Imbens (2016): ``Matching on the Estimated Propensity Score". 
\textit{Econometrica}, 84: 781-807.

Andrews, D.W.K.\ (1994): ``Empirical Process Methods in Econometrics,'' in Handbook of Econometrics, vol.\ IV, eds.\ R.F.\ Engle and D.L.\ McFadden, Elsevier. 

Andrews, D.\ W.\ K.\ and G.\ Soares (2010): ``Inference for
Parameters
Defined by Moment Inequalities Using Generalized Moment Selection". \textit{Econometrica}, 78: 119-157. 

Andrews, D.\ W.\ K.\ and X.\ Shi (2013): ``Inference Based on
Conditional Moment Inequalities". \textit{Econometrica}, 81: 609-666.

Anjali D.N.\ and P. Bossaerts (2014): ``Risk and Reward Preferences under Time Pressure''. \textit{Review of Finance}, 18: 999-1022.

Bamber, D.\ (1975): ``The Area above the Ordinal Dominance Graph and the Area below the Receiver Operating Characteristic Graph''. \textit{Journal of Mathematical Psychology} 12: 387-415.

Barrett, G.F.\ and  S.G.\ Donald (2003): ``Consistent Tests for Stochastic Dominance''. \textit{Econometrica}, 71:  71-104.

Bazzi, S., R.A.\ Blair, C.\ Blattman, O.\ Dube, M.\ Gudgeon and R.\ Peck (2021): ``The Promise and Pitfalls of Conflict Prediction: Evidence from Colombia and Indonesia''. \textit{The Review of Economics and Statistics}, forthcoming.  

Berge, T.J. and O.\ Jorda (2011): ``Evaluating the Classification of Economic Activity into Recessions and Expansions''. \textit{American Economic Journal: Macroeconomics} 3: 246-247.

Bonfim, D., G. Nogueira and S.\ Ongena (2021): `` `Sorry, We're Closed' Bank Branch Closures, Loan Pricing, and Information Asymmetries''. \textit{Review of Finance}, 25: 1211-1259.

%Carole, C-F.\ and T.J.\ Putnins (2014): ``Stock Price Manipulation: Prevalence and Determinants''. \textit{Review of Finance} 18: 23-66.

Clark, T.E.\ and M.W.\ McCracken (2012): ``In-sample Tests of Predictive Ability: A New Approach''. \textit{Journal of Econometrics} 170: 1-14. 

DeLong, E.R., D.M. DeLong and D.L. Clarke-Pearson (1988): ``Comparing Areas under Two or More Correlated Receiver Operating Characteristic Curves: A Nonparametric Approach''. \textit{Biometrics} 44: 837-845.

Demler, O.V., M.J.\ Pencina and R.B.\ D'Agostino, Sr. (2012): ``Misuse of DeLong Test to Compare AUCs for Nested Models''. \textit{Statistics in Medicine} 31: 2577-2587.

Egan, J.P.\ (1975): \emph{Signal Detection Theory and ROC Analysis}. Academic Press: New York.

Donald, S.G.\ and Y.-C.\ Hsu (2014): ``Estimation and Inference for Distribution Functions and Quantile Functions in Treatment Effect Models". \textit{Journal of Econometrics}, 178: 383-397.

Donald, S.G.\ and Y.-C.\ Hsu (2016): ``Improving the Power of Tests of Stochastic Dominance". \textit{Econometric Reviews}, 35: 553-585.

Donald, S.G.\ , Y.-C.\ Hsu and G.F.\ Barrett (2012): ``Incorporating Covariates in the Measurement of Welfare and Inequality: Methods and Applications". \textit{Econometrics Journal}, 15: C1-C30.

Elliott, G.\ and R.P.\ Lieli (2013): ``Predicting Binary Outcomes''. \textit{Journal of Econometrics}, 174: 15-26.

Hansen, P.\ R.\ (2005): ``A Test for Superior Predictive Ability".
\textit{Journal of Business and Economic Statistics}, 23: 365--380. 

Hsieh, F.\ and Turnbull, B.W.\ (1996): ``Nonparametric and Semiparametric Estimation of the Receiver Operating Characteristic Curve''. \textit{The Annals of Statistics}, 24: 25-40.

Inoue, A.\ and Kilian, L.\ (2004): ``In-sample or out-of-sample tests of predictability? Which one should we use?''. \textit{Econometric Reviews}, 23: 371-402.

Kleinberg, J., H.\  Lakkaraju, J.\ Leskovec, J.\ Ludwig and S.\ Mullainathan (2018): ``Human Decisions and Machine Predictions''. \textit{The Quarterly Journal of Economics}, 133: 237–293.

Lahiri, K.\ and L.\ Yang (2018): ``Confidence Bands for ROC Curves With Serially Dependent Data''. \textit{Journal of Business and Economic Statistics}, 36: 115-130.

Lahiri, K. and J.G. Wang (2013): ``Evaluating Probability Forecasts for GDP Declines Using Alternative Methodologies''. \textit{International Journal of Forecasting}, 29: 175-190.

Lieli, R.P.\ and Y-C.\ Hsu (2019): ``Using the Estimated AUC to Test the Adequacy of Binary Predictors''. \textit{Journal of Nonparametric Statistics}, 31: 100-130. 

Lieli, R.P.\ and A.\ Nieto-Barthaburu (2010): ``Optimal Binary Prediction for Group Decision Making''. \textit{Journal of Business and Economic Statistics}, 28: 308-319. 

Linton, O., E.\ Maasoumi and Y.-J.\ Whang (2005): ``Consistent Testing for Stochastic Dominance under General Sampling
Schemes". \textit{The Review of Economic Studies}, 72: 735-765. 

Linton, O., K.\ Song and Y.-J.\ Whang (2010): ``An Improved
Bootstrap Test of Stochastic Dominance". \textit{Journal of
Econometrics}, 154: 186-202. 

Ma, S.\ and M.R.\ Kosorok (2005): ``Robust Semiparametric M-estimation and the Weighted Bootstrap''. \textit{Journal of Multivariate Analysis}, 96: 190-270. 

McCracken, M.W., J.T.\ McGillicuddy and M.T.\ Owyang (2021): ``Binary Conditional Forecasts,'' \textit{Journal of Business and Economic Statistics}, forthcoming.

Pagan, A.\ (1984): ``Econometric Issues in the Analysis of Regressions with Generated Regressors''. \textit{International Economic Review}, 25: 221–247. 

Pepe, M.S.\ (2003): \emph{The Statistical Evaluation of Medical Tests for Classification and Prediction.} Oxford University Press: Oxford. 

Pollard, D.\ (1990): \emph{Empirical Processes: Theory and Application.}
CBMS Conference Series in Probability and Statistics, Vol.\ 2. Hayward, CA:
Institute of Mathematical Statistics. 

Schularik, M. and A.M. Taylor (2012): ``Credit Booms Gone Bust: Monetary Policy, Leverage Cycles, and Financial Crises, 1870-2008''. \textit{American Economic Review} 102: 1029-1061.

%Tantri, P. (2021): ``Fintech for the Poor: Financial Intermediation Without Discrimination,'' \textit{Review of Finance}, 25, pp.\ 561–593.

Van der Vaart, A.\ W.\ and J.\ A.\ Wellner (1996): \emph{Weak
Convergence and Empirical Processes: With Application to Statistics.}
New York: Springer-Verlag. 

Wooldridge, J.M.\ (2002): \emph{Econometric Analysis of Cross Section and Panel Data.} The MIT Press: Cambridge.

\end{apabib}

\newpage

\appendix

\begin{center}
{\Large Appendix}
\par\end{center}

\section{Auxiliary technical lemmas}\label{app: lemmas}

{\footnotesize

\begin{applemma}\label{lm: perturb}
\text{\rm [Stated in generic notation]} Let $X$ and $Y$ be random variables such that: (i) $|X|\le M$ a.s.\ for some $M>0$; and (ii) the density $f_Y$ of $Y$ exists and $f_Y\le C$ for some $C>0$. Then 
\[
\sup_{a\in \reals} |P(Y+hX\le a)-P(Y\le a)|\le CMh
\]
for all $h>0$ sufficiently small. 

\end{applemma}

\prf
As $|X|\le M$, we can write 
\begin{align*}
&P(Y\le a-hM)\le P(Y\le a-hX)\le P(Y\le a+hM)\\
\Leftrightarrow &F_Y(a-hM)-F_Y(a)\le P(Y\le a-hX)-P(Y\le a)\le F_Y(a+hM)-F_Y(a),
\end{align*} 
where $F_Y$ is the cdf of $Y$. Using the mean value theorem to expand the lower and upper bounds in the second inequality yields
\[
-f_Y(a-\lambda hM)hM\le P(Y\le a-hX)-P(Y\le a)\le f_Y(a+\theta hM)hM,
\]
where $\theta,\lambda\in[0,1]$. Given $f_Y\le C$, 
\[
|P(Y+hX\le a)-P(Y\le a)|\le hCM,
\]
where the upper bound does not depend on $a$.\eprf

\begin{applemma}\label{lm: cond indep}
\text{\rm [Stated in generic notation]} Let $Y\in\{0,1\}$ be a binary random variable and $X$ a random vector. Let $p(X)=P(Y=1|X)$. Then $Y$ and $X$ are independent conditional on $p(X)$.
\end{applemma}

\prf Let $f$ and $g$ be two bounded, continuous functions from $\reals$ to $\reals$. We need to show that the conditional expectation $E[f(X)g(Y)|p(X)]$ factors. By the law of iterated expectations,
\begin{equation}\label{eq: cond indep}
E[f(X)g(Y)|p(X)]=E\big\{E[f(X)g(Y)|X]\big|\, p(X)\big\}=E\big\{f(X)E[g(Y)|X]\big|\, p(X)\big\},
\end{equation}
where $E[g(Y)|X]=g(1)p(X)+g(0)[1-p(X)]$. This shows that $E[g(Y)|X]=E[g(Y)|p(X)]$. Substituting back into (\ref{eq: cond indep}),
\[
E[f(X)g(Y)|p(X)]=E\big\{f(X)E[g(Y)|p(X)]\big|\, p(X)\big\}=E[g(Y)\big|\,p(X)]E\big\{f(X)\big|\, p(X)\big\},
\]
which shows the claimed conditional independence.\eprf

\begin{applemma}\label{lm: qunatile unif conv}
\text{\rm [Stated in generic notation]}
Let $f(x)$ be a density supported on $[a,b]$ with $f(x)\ge m>0$ for all $x\in[a,b]$. Let $F$ be the corresponding cdf and $F^{-1}$ the corresponding quantile function. Let $\hat F_n$ be a sequence of non-decreasing random functions (such as the empirical cdf) satisfying $\sup_{x\in[a,b]}|\hat F_n(x)-F(x)|\rar_p 0$, and define $\hat F^{-1}_n(y)=\inf\{x: \hat F_n(x)\ge y\}$. Then:
\[
\sup_{y\in[0,1]}|\hat F^{-1}_n(y)-F^{-1}(y)|\rar_{p} 0.
\]
\end{applemma}

\prf The argument is lengthy and technical but entirely standard in the literature. It is omitted for brevity.\eprf

\begin{applemma}\label{lm: FP unif decomp}
Suppose that Assumptions \ref{assn: iid}, \ref{assn: beta-est}, \ref{assn: empirical process}, \ref{assn: cond densities} and \ref{assn: gradient unif} are satisfied. Then:

(i) $\widehat{FP}(c,\hat\beta)-FP(c,\hat\beta)=\widehat{FP}(c,\beta^*)-FP(c,\beta^*)+A_{1n}(c)$ with $\sup_{c\in[a_0,b_0]}|A_{1n}(c)|=o_p(n^{-1/2})$

(ii) $FP(c,\hat\beta)-FP(c,\beta^*)=\nabla_\beta\FP(c,\beta^*)(\hat\beta-\beta^*)+A_{2n}(c)$ with $\sup_{c\in[a_0,b_0]}|A_{2n}(c)|=o_p(n^{-1/2})$.

\end{applemma}

\prf (i) Let $\nu_n(c,\beta)=\sqrt{n}[\widehat{\FP}(c,\beta)-\FP(c,\beta)]$. Define $\delta_n=\|\hat\beta_n-\beta^*\|\ge 0$ and note that $\delta_n\rar_p 0$ by Assumption~\ref{assn: beta-est}. We can bound $\sqrt{n}A_{1n}(c)=\nu_n(c,\hat\beta)-\nu_n(c,\beta^*)$ as 
\begin{equation}\label{eq: A1 upper bd}
\sup_{c\in[a_0,b_0]}|\sqrt{n}A_{1n}(c)|\le
\sup_c|\nu_n(c,\hat\beta_n)-\nu_n(c,\beta^*)|\le \sup_{|c-c'|\le\delta_n,\, \|\beta-\beta'\|\le \delta_n}|\nu_n(c,\beta)-\nu_n(c',\beta')|.
\end{equation}
By Assumption~\ref{assn: empirical process}, the process $\nu_n(c,\beta)$ is stochastically equicontinuous w.r.t.\ $(c,\beta)$, which means that for any sequence of positive \emph{constants} $\delta_n\rar 0$, the last upper bound in (\ref{eq: A1 upper bd}) is $o_p(1)$. It is not hard to show that this remains true even when $\delta_n\rar_p 0$, but we omit this purely technical detail here (available on request).

%We complete the proof of part (i) by showing that this remains true even when $\delta_n\rar_p 0$ as here. Abbreviate the empirical process as $\nu_n(s)$, $s=(c,\beta)$, and set $d(s_1,s_2)=\max\{|c_1-c_2|, \|\beta_1-\beta_2\|\}$. For any $\epsilon,\eta>0$, there exists some $n_0\in\mathbb{N}$ such that 
%\begin{equation}\label{eq: stoch eq stoch conv pf}
%    P\Big[ \sup_{d(s_1,s_2)\le 1/n} |\nu_n(s_1)-\nu_n(s_2)|>\epsilon\Big]<\eta\text{ for all }n\ge n_0.
%\end{equation}
%Furthermore, 
%\begin{eqnarray*}
%&&P\Big[ \sup_{d(s_1,s_2)\le \delta_n} |\nu_n(s_1)-\nu_n(s_2)|>\epsilon\Big]=P\Big[ \sup_{d(s_1,s_2)\le \delta_n} |\nu_n(s_1)-\nu_n(s_2)|>\epsilon, \delta_n\le 1/n_0\Big]\\
%&&\qquad+P\Big[ \sup_{d(s_1,s_2)\le \delta_n} |\nu_n(s_1)-\nu_n(s_2)|>\epsilon, \delta_n> 1/n_0\Big]\\
%&&\le P\Big[ \sup_{d(s_1,s_2)\le 1/n_0} |\nu_n(s_1)-\nu_n(s_2)|>\epsilon, \delta_n\le 1/n_0\Big]+P\Big[\delta_n> 1/n_0\Big]\\
%&&\le P\Big[ \sup_{d(s_1,s_2)\le 1/n_0} |\nu_n(s_1)-\nu_n(s_2)|>\epsilon\Big]+P\Big[\delta_n> 1/n_0\Big]<\eta+o(1),
%\end{eqnarray*}
%where the last inequality follows from (\ref{eq: stoch eq stoch conv pf}) and $\delta_n\rar_p 0$. As $\eta$ is arbitrarily small, the proof is complete.

(ii) Using a mean value expansion, we can write
\begin{eqnarray*}
FP(c,\hat\beta)-FP(c,\beta^*)&=&\nabla_\beta\FP(c,\beta^*)(\hat\beta-\beta^*)+A_{2n}(c),
\end{eqnarray*}
where $A_{2n}(c)=[\nabla_\beta\FP(c,\ddot\beta_c)-\nabla_\beta\FP(c,\beta^*)](\hat\beta-\beta^*)$ and $\ddot\beta_c$ is on the line segment between $\hat\beta$ and $\beta^*$ for all $c$. Let $\delta_n=\|\hat\beta_n-\beta^*\|=o_p(1)$ and note that $\sup_c\|\hat\beta_n-\ddot\beta_c\|\le\delta_n$. Using the Cauchy-Schwarz inequality, we can bound $\sqrt{n}A_{2n}$ as 
\begin{align*}
\sup_{c\in [a_0,b_0]}|\sqrt{n}A_{2n}(c)|&\le\sup_c\big\|\nabla_\beta\FP(c,\ddot\beta_c) - \nabla_\beta\FP(c,\beta^*)\big\|\cdot\sqrt{n}\|\hat\beta-\beta^*\|\\
&\le\sup_{c\in[a_0,b_0],\,\beta\in B^*(\delta_n)} \big\|\nabla_\beta\FP(c,\beta)- \nabla_\beta\FP(c,\beta^*)\big\|\cdot\sqrt{n}\|\hat\beta-\beta^*\|=o_p(1)O_p(1)=o_p(1),
\end{align*}
where $\sqrt{n}\|\hat\beta-\beta^*\|=O_p(1)$ by Assumption~\ref{assn: beta-est}, and the supremum is $o_p(1)$, because Assumption \ref{assn: gradient unif} implies that $\nabla_\beta\FP(c,\beta)$ is uniformly continuous over $[a_0,b_0]\times B^*(s)$ for some $s>0$. \eprf

\begin{applemma}\label{lm: hatct to cstart}
Suppose that Assumptions \ref{assn: iid}, \ref{assn: beta-est}, \ref{assn: cond densities} and \ref{assn: gradient} are satisfied. Then: 
\[
\sup_{t\in[0,1]}|\hat c_t-c^*_t|=\sup_{t\in[0,1]}|\widehat{\FP}^{-1}_{\hat\beta}(t)-\FP^{-1}_{\beta^*}(t)|=o_p(1).
\]
\end{applemma}

\prf Using Lemma \ref{lm: FP unif decomp}, we write 
\begin{eqnarray}%\label{eq: FP decomp pf}
&&\widehat{\FP}(c,\hat\beta)-\FP(c,\beta^*)=\widehat{\FP}(c,\beta^*)-\FP(c,\beta^*)+\nabla_\beta\FP(\beta^*,c)(\hat\beta-\beta^*)+A_n(c),\nonumber
\end{eqnarray}
where $\sup_{c\in [a_0,b_0]}|A_n(c)|=o_p(1)$. Therefore,
\begin{eqnarray}
&&\sup_{c\in[a_0,b_0]}|\widehat{\FP}(c,\hat\beta)-\FP(c,\beta^*)|\le \sup_{c}|\widehat{\FP}(c,\beta^*)-\FP(c,\beta^*)|+\sup_{c}|\nabla_\beta\FP(\beta^*,c)(\hat\beta-\beta^*)|+o_p(1)\nonumber\\
&&\qquad\le \sup_{c}|\widehat{\FP}(c,\beta^*)-\FP(c,\beta^*)|+\sup_{c}\|\nabla_\beta\FP(\beta^*,c)\|\cdot\|\hat\beta-\beta^*\|+o_p(1).\label{eq: FP decomp pf}
\end{eqnarray}
Note that $1-\FP(c,\beta^*)$ is the cdf of $G(X,\beta^*)$ given $Y=0$ and $1-\widehat{\FP}(c,\beta^*)$ is the corresponding empirical cdf. Hence, $\sup_{c}|\widehat{\FP}(c,\beta^*)-\FP(c,\beta^*)|=o_p(1)$ by the Glivenko-Cantelli theorem. Furthermore, the second term in (\ref{eq: FP decomp pf}) is $o_p(1)$ as well, since $\sup_{c}\|\nabla_\beta\FP(\beta^*,c)\|\le M$ by Assumption \ref{assn: gradient unif} and $\|\hat\beta-\beta^*\|=o_p(1)$ by Assumption \ref{assn: beta-est}.  

Thus, we have shown that $\widehat{F}(c)\equiv 1-\widehat{\FP}(c,\hat\beta)$ is a (non-decreasing) random function that converges uniformly to the cdf $F(c)\equiv 1-\FP(c,\beta^*)$. The associated density is $f^*_0$, which is  bounded away from zero on $[a_0,b_0]$ by Assumption~\ref{assn: cond densities}(i). Therefore, we can apply Lemma~\ref{lm: qunatile unif conv} to conclude  
$\sup_{t\in [0,1]}|\widehat{F}^{-1}(t)-F^{-1}(t)|=o_p(1)\;\Leftrightarrow\;
\sup_{t\in [0,1]}|\widehat{\FP}^{-1}_{\hat\beta}(t)-\FP^{-1}_{\beta^*}(t)|=o_p(1)$,
given that $F^{-1}(t)=\FP^{-1}_{\beta^*}(1-t)$ and $\widehat{F}^{-1}(t)=\widehat{FP}^{-1}_{\hat\beta}(1-t)$. \eprf

\begin{applemma}\label{lm: uniform quantile}
\text{\rm [Stated in generic notation]} Let $X$ be a continuous random variable such that (i) the support of $X$ is a compact and connected set $[a,b]$;  and (ii) the density $f$ is continuous on $[a,b]$ with 
$f_x(c)\geq M\cdot \min\{c-a, b-c \}^K$ for all $c\in(a,b)$ and for some finite natural number $K$.  Then 
we have $\sup_{t\in[0,1]}|\widehat{c}_t-{c}_t|{\rightarrow}_p0$ where $c_t$ is the $t$-th quantile of $X$ and $\hat{c}_t$ is the estimator for $c_t$. 
\end{applemma}

\prf
We show the case in which $K=1$ and the proof for other cases is similar. 
Let $\widehat{F}(\cdot)$ the be estimator for the distribution function of $X$. Then
$\sup_{c\in[a,b]}|\widehat{F}(c)-F(c)|=O_P(n^{-1/2})$ where $F(\cdot)$ denotes  
the distribution function of $X$.
Another thing is that if $|\widehat{F}(\cdot)-F(\cdot)|=h$ where $h$ is small and if $f( c)\geq m>0$ on $[a^*,b^*]$, then
for all $\tau^*$ such that $a^*+ h/m \leq c_{\tau^*}\leq b^*- h/m$, we have $|\hat{c}_{\tau^*}-c_{\tau^*}|\leq h$.  To show this, note that if  $f( c)\geq m>0$ on $[a^*,b^*]$, then $|F(c_1)-F(c_2)|\geq |c_1-c_2|\cdot m$ for all $c_1,c_2\in [a^*,b^*]$.  Then it follows that for all $\tilde{c}\geq h/m$,
\begin{align*}
\widehat{F}(c_{\tau^*}+\tilde{c})&\geq
\widehat{F}(c_{\tau^*}+\tilde{c})-{F}(c_{\tau^*}+\tilde{c})+ {F}(c_{\tau^*}+\tilde{c})-{F}(c_{\tau^*})+\tau^*\\
&\geq - h + \tilde{c}*m +\tau^*\geq - h + h/m \cdot m+\tau^*=\tau^*.
\end{align*}
Similarly, it is true that  
$\widehat{F}(c_{\tau^*}+\tilde{c})\leq \tau^*$.  These two inequalities together imply that for all $\tau^*$ such that $a^*+ h/m \leq c_{\tau^*}\leq b^*- h/m$, we have
\begin{align*}
|\widehat{c}_{\tau^*}- c_{\tau^*} |\leq h/m.
\end{align*}
Under condition $f(c)\geq M\cdot \min\{c-a, b-c \}$, we pick $\ell_n\rightarrow 0$ with $(M\cdot \ell_n )^{-1}=o(n^{1/2})$.  This implies that the $f(c)\geq M\cdot \ell_n $ for all $c\in[a+\ell_n,b-\ell_n]$. It follows that
for all $\tau$ such that  $a+ 2\ell_n \leq c_{\tau}\leq b- 2\ell_n$, we have
\begin{align*}
|\widehat{c}_{\tau^*}- {c}_{\tau^*} | \leq 
\sup_{c\in[a,b]}|\widehat{F}(c)-F(c)| \cdot (M\cdot \ell_n )^{-1}= O_p(n^{-1/2})\cdot o (n^{1/2})=o_p(1).
\end{align*}
For $\tau$ such that $a\leq c_{\tau} < a+ 2\ell_n$, we have 
\begin{align*}
&~~a \leq \widehat{c}_{\tau} \leq{c}_{\tau}+|\widehat{F}(\cdot)-F(\cdot)| \cdot (M\cdot \ell_n )^{-1}\\
\Rightarrow &~~
a- c_\tau\leq \widehat{Q}_\tau-c_\tau \leq |\widehat{F}(\cdot)-F(\cdot)| \cdot (M\cdot \ell_n )^{-1}\\
\Rightarrow &~~
 |\widehat{c}_\tau-{c}_\tau|\leq \max\{
 2\ell_n,  |\widehat{F}(\cdot)-F(\cdot)| \cdot (M\cdot \ell_n )^{-1} \}=o_p(1).
\end{align*}
Similar result holds for $\tau$ such that $b-2\ell<  c_\tau \leq b$.  
Then these imply that
$\sup_{\tau\in[0,1]}|\widehat{c}_\tau-{c}_\tau|\stackrel{p}{\rightarrow}0$.\eprf
}%END of FOOTNOTESIZE

\section{Proofs of propositions and lemmas in the main text}\label{app: pf of results}

{\footnotesize

\paragraph{Proof of Proposition~\ref{prop: gradient \TP}} Write $TP(c,\beta)=1-P_1[G(X,\beta)\le c]$, where $P_1$ denotes probability conditional on $Y=1$. Let $e_j$ denote the $j$th unit vector with the same dimension as $\beta$. A second order Taylor expansion gives
\begin{eqnarray*}
G(X,\beta^*+he_j)=G(X,\beta^*)+G_j(X,\beta^*)h+G_{jj}(X,\beta^*+\lambda he_j)h^2,
\end{eqnarray*} 
where $\lambda\in[0,1]$. Take any $h_n\rar 0$. We want to compute the limit $\Delta$ of 
\begin{eqnarray*}
\Delta_n=\frac{1}{h_n}\Big\{ P_1\big[G(X,\beta^*+h_ne_j)\le c\big]-P_1\big[G(X,\beta^*)\le c\big]\Big\},
\end{eqnarray*}
as $n\rar\infty$. Using the Taylor expansion above, we can write
\begin{eqnarray*}
\Delta_n=\frac{1}{h_n}\Big\{ P_1\big[G^*+G^*_jh_n+G_{jj}^{(n)}h_n^2\le c\big]-P_1\big[G^*\le c\big]\Big\},
\end{eqnarray*}
where $G^*=G(X,\beta^*)$, $G^*_j=G_j(X,\beta^*)$ and $G_{jj}^{(n)}=\partial_{jj}G(X,\beta^*+\lambda_n h_n e_j)$ with $\lambda_n\in[0,1]$.
Using the law of iterated expectations,
\begin{eqnarray}
\Delta_n&=&E_1\left\{ \frac{P_1\big[G^*+G_{jj}^{(n)}h_n^2\le c-G^*_jh_n\mid G_j^*\big]-P_1\big[G^*\le c\mid G^*_j\big]}{h_n}\right\}\nonumber\\
%&=&-E_1\left\{G_j^*\times\frac{P_1\big[G^*+G_{jj}^{(n)}h_n^2\le c-G^*_jh_n\mid G_j^*\big]-P_1\big[G^*\le c\mid G^*_j\big]}{-G_j^* h_n}\right\}\\
&=&E_1\left\{\frac{P_1\big[G^*\le c-G^*_jh_n\mid G_j^*\big]-P_1\big[G^*\le c\mid G^*_j\big]}{h_n}\right.\nonumber\\
&+&\left.\frac{P_1\big[G^*+G_{jj}^{(n)}h_n^2\le c-G^*_jh_n\mid G_j^*\big]-P_1\big[G^*\le c-G^*_jh_n\mid G_j^*\big]}{h_n}\right\}
\label{derivative O(h2) term}
\end{eqnarray}
where $E_1$ is expectations w.r.t.\ $P_1$.  By Assumptions~\ref{assn: gradient}(ii) and \ref{assn: gradient}(iii), $G_{jj}^{(n)}$ is a bounded random variable, and the conditional density of $G^*$ given $G^*_j$ (and $Y=1$) exists, and is also bounded, uniformly in $G^*_j$. Thus, applying Lemma~\ref{lm: perturb} gives 
\begin{eqnarray}
&&\sup_a\Big|P_1\big[G^*+G_{jj}^{(n)}h_n^2\le a\mid G_j^*\big]-P_1\big[G^*\le a\mid G_j^*\big]\Big|\le Kh_n^2\label{identical asy dist}
\end{eqnarray}
for some $K>0$. This inequality, in turn, implies that the second term within the expectation in equation (\ref{derivative O(h2) term}) is $O(h_n)$. Therefore, we can write $\Delta_n$ as
\begin{eqnarray*}
\Delta_n
&=&-E_1\left\{G^*_j\times\frac{P_1\big[G^*\le c-G_j^* h_n\mid G_j^*\big]-P_1\big[G^*\le c\mid G_j^*\big]}{-G_j^* h_n}+O(h_n)\right\}\\
&=&-E_1\left\{G^*_j\times f_{G^*|G^*_j}(c-\theta G^*_j h_n\mid G_j^*)+O(h_n)\right\},
\end{eqnarray*}
where the second equality follows from the mean value theorem with $f_{G^*|G^*_j}$ denoting the conditional density of $G^*$ given $G^*_j$ (and $Y=1$) and $\theta\in[0,1]$. 

Now, inequality (\ref{identical asy dist}) shows that the error term $O(h_n)$ is dominated in absolute value by a constant multiple of $h_n$. Furthermore, by Assumptions~\ref{assn: gradient}(iii) and \ref{assn: gradient}(iv), $f_{G^*|G^*_j}$ is bounded uniformly in $G_j^*$ and $E_1|G_j^*|<\infty$. This allows us to apply the dominated convergence theorem to conclude
\begin{eqnarray*}
\Delta&=&\lim_{n\rar \infty}\Delta_n=-E_1\left\{G_j^*\times f_{G^*|G^*_j}(c\mid G_j)\right\}.
\end{eqnarray*}
Finally, note that
\begin{eqnarray*}
\Delta&=&-E_1\left\{G_j^*\times f_{G^*|G^*_j}(c\mid G^*_j)\right\}\\
&=&-\int t f_{G^*|G^*_j}(c|t)f_{G_j^*}(t)dt=-\int t f_{G^*_j|G^*}(t|c)f_{G^*}(c)dt\\
&=& -E_1\big[G_j^*|G^*=c\big]f_{G^*}(c),
\end{eqnarray*}
where $f_{G^*_j|G^*}$ is the conditional density of $G_j^*$ given $G^*$ (and $Y=1$) and $f_{G^*}$ is the density of $G^*$ (given $Y=1$). The last expression is equivalent to equation (\ref{eq: gradient \TP}) in Proposition~\ref{prop: gradient \TP}.

The the second part of Proposition~\ref{prop: gradient \TP} follows immediately from Lemma~\ref{lm: cond indep} and observing that under correct specification $G(X,\beta^*)=G(X,\beta^\circ)=p(X)$.\eprf

\paragraph{Proof of Proposition~\ref{prop: logit}} Setting $G(X,\beta)=\Lambda(\tilde X'\beta)$, where $\tilde X=(1,X')'$ and $\beta=(\beta_0,\beta_1,\ldots,b_k)'$, it is straightforward to verify that 
\[
\frac{\partial}{\partial\beta_j}\Lambda(\tilde X'\beta)=\Lambda(\tilde X'\beta)[1-\Lambda(\tilde X'\beta)]X_j,\quad j=0,1,\ldots, k
\]
with $X_0\equiv 1$. Taking expectations conditional on $\Lambda(\tilde X'\beta)=c$ and $Y=1$ gives
\[
E\left[\frac{\partial}{\partial\beta_j}\Lambda(\tilde X'\beta)\Big|\,\Lambda(\tilde X'\beta)=c, Y=1\right]=c(1-c)E[X_j|\,\Lambda(\tilde X'\beta)=c, Y=1]
\]
According to the general formula (\ref{eq: gradient \TP}), multiplying by $f_1^*(c)=f_\Lambda(c|Y=1)$ gives the $j$th component of $\nabla_\beta\TP(c,\beta)$.\eprf

%\begin{applemma}\label{lm: unif R}
%Consider the expansions given in equations (\ref{eq: R1}), (\ref{eq: R2}) and (\ref{eq: R3}) and suppose that Assumptions \ref{assn: iid} through \ref{assn: gradient} are satisfied. Then the remainder terms converge to zero uniformly as follows: (i) $\sup_{t\in [0,1]} R_{1n}(t)=o_p(1)$; (ii) $\sup_{t\in [0,1]} R_{2n}(t)=o_p(1)$; (iii) $\sup_{t\in I_\epsilon}R_{3n}(t)=o_p(1)$, where $I_\epsilon=\{t: c_t\in[a_1+\epsilon, b_1-\epsilon]\}=[\FP(b_1-\epsilon, \beta^*), \FP(a_1+\epsilon,\beta^*)]$. %(iv) $\sup_{t\in (0,1)} R_{4n}(t)=o_p(1)$. 
%\end{applemma}

\paragraph{Proof of Lemma~\ref{lm: FP delta meth}} 
%The proof is similar to the proof of Lemma~\ref{lm: FP unif decomp}.
(i) Let $\nu_n(c,\beta)=\sqrt{n}[\widehat{\TP}(c,\beta)-\TP(c,\beta)]$. We can then write $R_{1n}(t)=\nu_n(\hat c_t,\hat\beta)-\nu_n(c^*_t,\beta^*)$. Define $\delta_n=\max\big\{\sup_t|\hat c_t-c^*_t|, \|\hat\beta_n-\beta^*\|\big\}$ and note that $\delta_n\rar_p 0$ by Lemma~\ref{lm: hatct to cstart} and Assumption~\ref{assn: beta-est}. We can bound $R_{1n}(t)$ as
\begin{equation}\label{eq: R1 upper bd}
\sup_t|R_{1n}(t)|=\sup_t|\nu_n(\hat c_t,\hat\beta)-\nu_n(c^*_t,\beta^*)|\le\sup_{|c-c'|\le\delta_n,\, \|\beta-\beta'\|\le \delta_n}|\nu_n(c,\beta)-\nu_n(c',\beta')|.
\end{equation}
By Assumption~\ref{assn: empirical process}, the process $\nu_n(c,\beta)$ is stochastically equicontinuous w.r.t.\ $(c,\beta)$, which means that for any sequence of positive constants $\delta_n\rar 0$, the r.h.s.\ of (\ref{eq: R1 upper bd}) is $o_p(1)$. It is not hard to show that this remains true even when $\delta_n\rar_p 0$, but we omit this purely technical detail here (available on request).

%Abbreviate the empirical process as $\nu_n(s)$, $s=(c,\beta)$, and set $d(s_1,s_2)=\max\{|c_1-c_2|, \|\beta_1-\beta_2\|\}$. For any $\epsilon,\eta>0$, there exists some $n_0\in\mathbb{N}$ such that 
%\begin{equation}\label{eq: stoch eq stoch conv pf}
%    P\Big[ \sup_{d(s_1,s_2)\le 1/n} |\nu_n(s_1)-\nu_n(s_2)|>\epsilon\Big]<\eta\text{ for all }n\ge n_0.
%\end{equation}
%Furthermore, 
%\begin{eqnarray*}
%&&P\Big[ \sup_{d(s_1,s_2)\le \delta_n} |\nu_n(s_1)-\nu_n(s_2)|>\epsilon\Big]=P\Big[ \sup_{d(s_1,s_2)\le \delta_n} |\nu_n(s_1)-\nu_n(s_2)|>\epsilon, \delta_n\le 1/n_0\Big]\\
%&&\qquad+P\Big[ \sup_{d(s_1,s_2)\le \delta_n} |\nu_n(s_1)-\nu_n(s_2)|>\epsilon, \delta_n> 1/n_0\Big]\\
%&&\le P\Big[ \sup_{d(s_1,s_2)\le 1/n_0} |\nu_n(s_1)-\nu_n(s_2)|>\epsilon, \delta_n\le 1/n_0\Big]+P\Big[\delta_n> 1/n_0\Big]\\
%&&\le P\Big[ \sup_{d(s_1,s_2)\le 1/n_0} |\nu_n(s_1)-\nu_n(s_2)|>\epsilon\Big]+P\Big[\delta_n> 1/n_0\Big]<\eta+o(1),
%\end{eqnarray*}
%where the last inequality follows from (\ref{eq: stoch eq stoch conv pf}) and $\delta_n\rar_p 0$. As $\eta$ is arbitrarily small, the proof is complete.

(ii) We can write 
$R_{2n}(t)=\big[\nabla_\beta\TP(\hat c_t,\ddot\beta_t)- \nabla_\beta\TP(c^*_t,\beta^*)\big]\sqrt{n}(\hat\beta-\beta^*)$,
where $\ddot\beta_t$ is on the line segment connecting $\hat\beta$ and $\beta^*$ for all $t$. Set $\delta_n=\max\big\{\sup_t|\hat c_t-c^*_t|, \|\hat\beta-\beta^*\|\big\}$ and note that $\delta_n\rar_p 0$ by Lemma~\ref{lm: hatct to cstart} and Assumption~\ref{assn: beta-est}. Furthermore, note that $\sup_t\|\ddot \beta_t-\beta^*\|\le\delta_n$. We can then bound $R_{2n}(t)$ as
\begin{align*}
\sup_{t\in[0,1]}|R_{2n}(t)|&\le\sup_t\big\|\nabla_\beta\TP(\hat c_t,\ddot\beta)- \nabla_\beta\TP(c^*_t,\beta^*)\big\|\sqrt{n}\|\hat\beta-\beta^*\|\\
&\sup_{|c-c'|\le\delta_n,\, \beta\in B^*(\delta_n)} \big\|\nabla_\beta\TP(c,\beta)- \nabla_\beta\TP(c',\beta^*)\big\|\le\sqrt{n}\|\hat\beta-\beta^*\|=o_p(1)O_p(1)=o_p(1),
\end{align*}
where $\sqrt{n}\|\hat\beta-\beta^*\|=O_p(1)$ by Assumption~\ref{assn: beta-est}, and the supremum is $o_p(1)$, because Assumption \ref{assn: gradient unif} implies that $\nabla_\beta\TP(c,\beta)$ is uniformly continuous over $[a_1,b_1]\times B^*(s)$ for some $s>0$. 

%\textcolor{red}{The last step is not properly justified! It's a type of uniform continuity condition on $\nabla_\beta TP$. What are the weakest primitive conditions to ensure this? Do we need to restrict $c$ to a compact set? Do we need to restrict $\beta$ to a closed ball around $\beta^*$?} \textcolor{red}{[ I think the weakest condition for uniform continuity of $\nabla_\beta TP$ is the absolute value of the derivatives are bounded above.  We may restrict $\beta$ to a closed ball because for parametric estimation of $\beta$, we would in general consider a compact subset that contains the true (pseudo-true) parameters of interest. ] }

(iii) We can write $R_{3n}(t)=\sqrt{n}(\hat c_t - c_t^*)[f^*_1(\ddot c_t)-f^*_1(c^*_t) ]$, where $\ddot c_t$ is on the line segment connecting $\hat c_t$ and $c^*_t$. Set $\delta_n=\sup_t|\hat c_t-c^*_t|$ and note that $\delta_n\rar_p 0$ by Lemma~\ref{lm: hatct to cstart}. We can bound $R_{3n}(t)$ over $T$ as 
\begin{align*}
\sup_{t\in T}|R_{3n}(t)|&\le \sqrt{n}\sup_t|\hat c_t-c^*_t|\sup_t|f_1^*(\ddot c_t)-f_1^*(c^*_t)|\\
&\le\sqrt{n}\sup_t|\hat c_t-c^*_t|\sup_{c,c'\in [c_{0,L}, c_{0,U}],\, |c-c'|\le\delta_n}|f_1^*(c)-f_1^*(c')|=O_p(1)o_p(1)=o_p(1),
\end{align*}
where $\sqrt{n}\sup_t|\hat c_t-c^*_t|=O_p(1)$ by the functional delta method (see part (v) below), and the supremum is $o_p(1)$, because Assumption~\ref{assn: cond densities} implies that $f^*_1(c)$ is uniformly continuous over the closed interval $[c_{0,L}, c_{0,U}]$. To see this, write $f_1^*=(f_1^*/f_0^*)\cdot f_0^*$. The ratio $f_1^*/f_0^*$ is continuous over $[c_{0,L}, c_{0,U}]$ by Assumption~\ref{assn: cond densities}(iii). The density $f_0^*$ is continuous over $[a_0,b_0]\supseteq [c_{0,L}, c_{0,U}]$ by Assumption~\ref{assn: cond densities}(iii). Therefore, $f_1^*$ is also continuous over the compact interval $[c_{0,L}, c_{0,U}]$, which means that it is uniformly continuous.

(iv) Using Lemma \ref{lm: FP unif decomp}, we write 
\begin{eqnarray}
&&\sqrt{n}[\widehat{\FP}(c,\hat\beta)-\FP(c,\beta^*)]=\sqrt{n}[\widehat{\FP}(c,\beta^*)-\FP(c,\beta^*)]+\nabla_\beta\FP(\beta^*,c)\sqrt{n}(\hat\beta-\beta^*)+A_n(c),\nonumber
\end{eqnarray}
where $\sup_{c\in[a_0,b_0]}|A_n(c)|=o_p(1)$. But
\begin{eqnarray*}
\sqrt{n}[\widehat{\FP}(c,\beta^*)-\FP(c,\beta^*)]&=&\frac{1}{\sqrt{n}}\sum_{i=1}^n \frac{1-Y_i}{1-\hat\pi}\big[1(G(X_i,\beta^*)>c)-\FP(c,\beta^*)\big]\\
&=&\frac{1}{\sqrt{n}}\sum_{i=1}^n \frac{1-Y_i}{1-\pi}\big[1(G(X_i,\beta^*)>c)-\FP(c,\beta^*)\big]+B_n(c),
\end{eqnarray*}
where 
\[
B_n(c)=\left(\frac{1}{1-\hat\pi}-\frac{1}{1-\pi}\right)\cdot \frac{1}{\sqrt{n}}\sum_{i=1}^n (1-Y_i)\big[1(G(X_i,\beta^*)>c)-\FP(c,\beta^*)\big].
\]
Clearly, $E\big\{(1-Y_i)[1(G(X_i,\beta^*)>c)-\FP(c,\beta^*)]\big\}=0$, so $\sup_c|B_n(c)|$ is of the form $o_p(1)\cdot O_p(1)=o_p(1)$. Furthermore, by Assumption~\ref{assn: beta-est},
\[
\sqrt{n}(\hat\beta-\beta^*)=\frac{1}{\sqrt{n}}\sum_{i=1}^n\psi_{\beta}(Y_i,X_i,\beta^*)+o_p(1),
\]
where the remainder term does not depend on $c$ at all. It follows that the asymptotically linear representation of $\widehat{\FP}(c,\hat\beta)$ holds uniformly over $c\in[a_0,b_0]$.

(v) By Assumption~\ref{assn: cond densities}, the function $c\mapsto \FP(c,\beta^*)$ is a survivor function with compact support $[a_0,b_0]$ and a continuous density $f^*_0(c)$ that is bounded away from zero on $[a_0,b_0]$. By, for example, Lemma 3.9.23 of van der Vaart and Wellner (1996), the inverse map $\phi(F)=F^{-1}$ is Hadamard differentiable at $\FP(\cdot,\beta^*)$ (tangentially to $C[a_0,b_0]$) and the Hadamard derivative is the map $\phi'_{FP}(h)=h(FP^{-1}_{\beta^*})/f_0^*(FP^{-1}_{\beta^*})$. The functional delta method (e.g., Theorem 3.9.4 of \emph{ibid.}) then gives 
\begin{eqnarray*}
\sqrt{n}[\widehat{FP}_{\hat\beta}^{-1}(t)- \FP^{-1}_{\beta^*}(t)]&=& \phi'_{FP}\Big(\sqrt{n}\big[\widehat{FP}\big(\cdot,\hat\beta\big)-\FP\big(\cdot,\beta^*\big)\big]\Big)+R_{5n}(t)\\
&=&\frac{1}{f_0^*(c^*_t)}\sqrt{n}[\widehat{FP}_{\hat\beta}(c^*_t)- \FP_{\beta^*}(c^*_t)]+R_{5n}(t),
\end{eqnarray*}
where $\sup_{t\in(0,1)}|R_{5n}(t)|=o_p(1)$ because $f_0^*$ is bounded away from zero on $[a_0,b_0]$ (see Example 3.9.24 of \emph{ibid.})\eprf

\paragraph{Proof of Proposition~\ref{prop: ROC curve asymptotics}} 

(i) The uniformity of the influence function representation for $\sqrt{n}[\widehat{R}(t,\hat\beta)-R(t,\beta^*)]$ follows from: equations  (\ref{eq: func decomp}) through (\ref{eq: R5}), the uniform asymptotic negligiblity results stated in Lemma~\ref{lm: FP delta meth}, and the fact that the influence function representation of $\sqrt{n}[\widehat{TP}(c,\beta^*)-\TP(c,\beta^*)]$ holds uniformly over $c\in[a_1,b_1]\supseteq [c_{0,L},c_{0,U}]$. The proof of this last fact is similar to the proof of Lemma~\ref{lm: FP delta meth}(iv) and is omitted. 

(ii)  We write $\psi_{R,n}(t)\equiv\frac{1}{\sqrt{n}}\sum_{i=1}^n  \psi_{R}(Y_i,X_i,t,\beta^*)$, $\psi_{TP,n}(c)\equiv\frac{1}{\sqrt{n}}\sum_{i=1}^n \psi_{TP}(Y_i,X_i,c,\beta^*)$, $\psi_{FP,n}(c)\equiv\frac{1}{\sqrt{n}}\sum_{i=1}^n \psi_{FP}(Y_i,X_i,c,\beta^*)$, and $\lambda(c)=f_1^*(c)/f_0^*(c)$ so that 
%\begin{align}
%&\psi_{R,n}(t)\equiv\frac{1}{\sqrt{n}}\sum_{i=1}^n  %&\psi_{TP,n}(c)\equiv\frac{1}{\sqrt{n}}\sum_{i=1}^n \psi_{TP}(Y_i,X_i,c,\beta^*),\nonumber\\
%&\psi_{FP,n}(c)\equiv\frac{1}{\sqrt{n}}\sum_{i=1}^n \psi_{FP}(Y_i,X_i,c,\beta^*)\nonumber
%\end{align}
\begin{equation*}
\psi_{R,n}(t)=\psi_{TP,n}(c_t^*)-\lambda(c^*_t)\psi_{FP,n}(c_t^*).    
\end{equation*}
Let $\delta_n>0$ be an arbitrary sequence with $\delta_n\rar 0$. We need to show that $\sup_{t,t'\in T,\, |t-t'|\le\delta_n}|\psi_{R,n}(t)-\psi_{R,n}(t')|\rar_p 0$ (see Andrews 1994 for various equivalent definitions of stochastic equicontinuity). 

First note that $|c^*_t-c^*_{t'}|\le M|t-t'|$ for some $M>0$ because $f^*_0$ is bounded away from zero on $[a_0,b_0]$. Hence, $|t-t'|\le\delta_n$ implies $|c^*_t-c^*_{t'}|\le M\delta_n\equiv\delta'_n$. Next we write
\begin{eqnarray*}
&&\psi_{R,n}(t)-\psi_{R,n}(t')=\psi_{TP,n}(c^*_t)-\psi_{TP,n}(c^*_{t'})
-\lambda(c^*_t)[\psi_{FP,n}(c^*_t)-\psi_{FP,n}(c^*_{t'})]-\psi_{FP,n}(c^*_{t'})[\lambda(c^*_t)-\lambda(c^*_{t'})]
\end{eqnarray*}
so that for $t,t'\in T$ and $c,c'\in [c_{0,L},c_{0,U}]$,
\begin{eqnarray*}
\sup_{|t-t'|\le\delta_n}|\psi_{R,n}(t)-\psi_{R,n}(t')|\le\sup_{|c-c'|\le\delta'_n}|\psi_{TP,n}(c)-\psi_{TP,n}(c')|&+&\sup_c|\lambda(c)|\sup_{|c-c'|\le\delta'_n}|\psi_{FP,n}(c)-\psi_{FP,n}(c')| \\
&-&\sup_c|\psi_{FP,n}(c)|\sup_{|c-c'|\le\delta'_n}[\lambda(c)-\lambda(c')|.
\end{eqnarray*}
The argument can be completed by showing the stochastic equicontinuity of the processes $\psi_{FP,n}(c)$ and $\psi_{TP,n}(c)$ and exploiting the uniform continuity of $\lambda(c)$ over $[c_{0,L}, c_{0,U}]$ (Assumption~\ref{assn: cond densities}(iii)). For illustration, here we show the stochastic equicontinuity of $\psi_{FP,n}(c)$.

Define
\[
\nu^*_n(c)=\frac{1}{\sqrt{n}}\sum_{i=1}^n \frac{1-Y_i}{1-\pi}\big[1(G(X_i,\beta^*)>c)-\FP(c,\beta^*)\big]
\]
Assumption \ref{assn: empirical process} implies that $\nu^*_n(c)$ is stochastically equicontinuous over $[a_0,b_0]$. We can write 
\begin{eqnarray}
&&\sup_{|c-c'|\le \delta'_n}\Big| \frac{1}{\sqrt{n}}\sum_{i=1}^n\psi_{FP}(Y_i,X_i,c,\beta^*)-\frac{1}{\sqrt{n}}\sum_{i=1}^n\psi_{FP}(Y_i,X_i,c',\beta^*) \Big|\le \sup_{|c-c'|\le \delta'_n}|\nu^*_n(c)-\nu^*_n(c')|\nonumber\\
&&\qquad\qquad\qquad\qquad+\sup_{|c-c'|\le \delta'_n}\|\nabla_\beta\FP(c,\beta^*)-\nabla_\beta\FP(c',\beta^*)\| \cdot\Big\|\frac{1}{\sqrt{n}}\sum_{i=1}^n\psi_{\beta}(Y_i,X_i,\beta^*)\Big\|\label{eq: psiFP stoch equic}
\end{eqnarray}
Then $\sup_{|c-c'|\le \delta'_n}|\nu^*_n(c)-\nu^*_n(c')|=o_p(1)$ by stochastic equicontinuity and $\sup_{|c-c'|\le \delta'_n}|\nabla_\beta\FP(c,\beta^*)-\nabla_\beta\FP(c',\beta^*)|=o_p(1)$ because $c\mapsto \nabla_\beta\FP(c,\beta^*)$ is uniformly continuous over $[a_0,b_0]$ (and hence over $[c_{0,L},c_{0,U}]$) by Assumption \ref{assn: gradient unif}. Finally, the central limit theorem implies $\|n^{-1/2}\sum_{i=1}^n\psi_{\beta}(Y_i,X_i,\beta^*)\|=O_p(1)$ under Assumptions~\ref{assn: iid} and \ref{assn: beta-est}. Hence, the r.h.s.\ of inequality (\ref{eq: psiFP stoch equic}) is of the form $o_p(1)+o_p(1)O_p(1)=o_p(1)$, which means that the process $c\mapsto \frac{1}{\sqrt{n}}\sum_{i=1}^n\psi_{FP}(Y_i,X_i,c,\beta^*)$ is stochastically equicontinuous. 

(iii) The multivariate central limit theorem implies that the finite dimensional projections of the process $t\mapsto \frac{1}{\sqrt{n}}\sum_{i=1}^n\psi_{R}(Y_i,X_i,t,\beta^*)$ converge in distribution to multivariate normal vectors with covariance matrices corresponding to $h_{R}$. Coupled with stochastic equicontinuity, this is sufficient (and necessary) for the weak convergence of the whole process in $L^\infty(T)$ to a Gaussian process with the given finite dimensional distributions; see, e.g., van der Vaart and Wellner (1996, Ch.\ 1.5). 
Finally, the process $t\mapsto\sqrt{n}[\widehat{R}(t,\hat\beta)-R(t,\beta^*)]$ has the same limit distribution because of part (i). \eprf

\paragraph{Proof of Proposition~\ref{prop: weighted bootstrap}}
The proof follows the same steps that we discussed after Lemma \ref{lm: FP delta meth}  to show Proposition \ref{prop: ROC curve asymptotics} and we omit the details.\eprf

\paragraph{Proof of Proposition~\ref{prop: multiplier bootstrap}}
We first claim that $\{\hat{\psi}_{TP}(Y_i,X_i,\hat{c}_{t},\hat\beta):t\in\mathcal{T}, 1\leq i\leq n, n\geq 1\}$ is manageable in the sense of Definition 7.9 of Pollard (1990). Note that 
$\{{Y_i}/\hat{\pi}\big(1[G(x,\beta)> c]-\widehat{\TP}(c,\beta)\big): c\in\mathcal{C}, 1\leq i\leq n, n\geq 1\}$ 
is a Type I class of functions as in Andrews (1994), so it is manageable w.r.t.\ 
$\{{2|Y_i|}/\hat{\pi}: 1\leq i\leq n, n\geq 1\}$. 
In addition, $\{\nabla_\beta \widehat{\TP}(c,\hat\beta)\hat{\psi}_\beta(Y_i,X_i,\hat{\beta}): c\in\mathcal{C}, 1\leq i\leq n, n\geq 1  \}$ is manageable w.r.t.\ $\{M|\hat{\psi}_\beta(Y_i,X_i,\hat{\beta})|: 1\leq i\leq n, n\geq 1  \}$ for some large $M>0$
because 
it is a Type II class of functions given that $\nabla_\beta \widehat{\TP}(c,\hat{\beta})$ is Lipschitz continuous in $c$ and bounded above by Assumption \ref{assn: consistent estimator}. Then by Theorem of Andrews (1994), $\{\hat{\psi}_{TP}(Y_i,X_i,\hat{c}_{t},\hat\beta):t\in\mathcal{T}, 1\leq i\leq n, n\geq 1\}$ is manageable
 w.r.t.\  $\{2/\hat{\pi}+M|\hat{\psi}_\beta(Y_i,X_i,\hat{\beta})|: 1\leq i\leq n, n\geq 1\}$.  Similarly, $\{\hat{\psi}_{FP}(Y_i,X_i,\hat{c}_{t},\hat\beta):t\in\mathcal{T}, 1\leq i\leq n, n\geq 1\}$ is manageable
  w.r.t.\  $\{{2}/(1-\hat{\pi})+M|\hat{\psi}_\beta(Y_i,X_i,\hat{\beta})|: 1\leq i\leq n, n\geq 1\}$.  By  Assumption \ref{assn: consistent estimator},  ${\hat{f}_{1}(c)}/{\hat{f}_{0}(c)}$ is Lipschitz continuous in $c$ and bounded above, so it is true that $\{\widehat{\psi}_R(Y_i,X_i,t,\hat\beta): t\in\mathcal{T},1\leq i\leq n, n\geq 1 \}$ is manageable.

Next, given Assumption \ref{assn: consistent estimator}, it is straightforward to see that 
\begin{align*}
\frac{1}{n}\sum_{i=1}^n \widehat{\psi}_R(Y_i,X_i,t_1,\hat\beta) \widehat{\psi}_R(Y_i,X_i,t_2,\hat\beta)
\stackrel{p}{\rightarrow} h_R(t_1,t_2)
\end{align*}
uniformly over $t_1,t_2\in\mathcal{T}$.  Then, these are sufficient to show Proposition~\ref{prop: weighted bootstrap}.\eprf

}%END OF FOOTNOTESIZE

\section{Uniformly consistent estimation of $\psi_R$}\label{app: uc est psiR}
{\footnotesize
We give estimators that satisfy Assumption \ref{assn: consistent estimator}.  We focus on the case where $t_\ell=0$ and $t_u=1$ so that $\mathcal{C}=[a_0,b_0]$ because the results for other cases are similar.   

Let $h$ denote a bandwidth that depends on sample size $n$ and
$K(u)$ a kernel function. %We first consider the estimation of ${\FP}(c,{\beta}^*)$ and ${\TP}(c,{\beta}^*)$
%because the proof for the density estimators is more involved. 

Define primary estimators $\nabla_\beta
\widetilde{\FP}(c,\hat{\beta})$, $\nabla_\beta
\widetilde{\TP}(c,\hat{\beta})$, $\tilde{f}_{1}(c)$ and $\tilde{f}_{0}(c)$ as
\begin{align*}
&\nabla_\beta\widetilde{\FP}(c,\hat{\beta})=\frac{1}{\hat{\pi}}\frac{1}{nh}\sum_{i=1}^n \nabla_\beta G(X_i,\hat{\beta})\cdot Y_i \cdot K\Big(\frac{G(X_i,\hat{\beta})-c}{h}\Big),\\
&\nabla_\beta\widetilde{TP}(c,\hat{\beta})=\frac{1}{1-\hat{\pi}}\frac{1}{nh}\sum_{i=1}^n \nabla_\beta G(X_i,\hat{\beta})\cdot (1-Y_i) \cdot K\Big(\frac{G(X_i,\hat{\beta})-c}{h}\Big),\\
&\tilde{f}_{1}(c)=\frac{1}{\hat{\pi}}\frac{1}{nh}\sum_{i=1}^n Y_i \cdot K\Big(\frac{G(X_i,\hat{\beta})-c}{h}\Big),\\
&\tilde{f}_{0}(c)=\frac{1}{1-\hat{\pi}}\frac{1}{nh}\sum_{i=1}^n (1-Y_i)\cdot K\Big(\frac{G(X_i,\hat{\beta})-c}{h}\Big).
\end{align*}
The final estimators are defined as
\begin{align}
&\nabla_\beta
\widehat{\FP}(c,\hat{\beta})=\left\{
\begin{array}{ll}
\nabla_\beta\widetilde{\FP}(a_{0}+h,\hat{\beta}) & \mbox{if $c\in [a_{0},a_{0}+h]$,} \\
\nabla_\beta\widetilde{\FP}(c,\hat{\beta}) & \mbox{if $c\in[a_{0}+h,b_0-h]$,} \\
\nabla_\beta\widetilde{\FP}(b_0-h,\hat{\beta}) & \mbox{if $c\in[b_0-h,b_0]$,}
\end{array}
\right. \nonumber\\
&\nabla_\beta\widehat{\TP}(c,\hat{\beta})=\left\{
\begin{array}{ll}
\nabla_\beta\widetilde{\TP}(a_{0}+h,\hat{\beta}) & \mbox{if $c\in [a_{0},a_{0}+h]$,} \\
\nabla_\beta\widetilde{\TP}(c,\hat{\beta}) & \mbox{if $c\in[a_{0}+h,b_0-h]$,} \\
\nabla_\beta\widetilde{\TP}(b_0-h,\hat{\beta}) & \mbox{if $c\in[b_0-h,b_0]$,}
\end{array}
\right. \nonumber
\\
&\hat{f}_{1}(c)=\left\{
\begin{array}{ll}
\tilde{f}_{1}(a_{0}+\delta_n) &\mbox{if $c\in [a_{0},a_{0}+\delta_n]$,} \\
\tilde{f}_{1}(c) & \mbox{if $c\in[a_{0}+\delta_n,b_0-\delta_n]$,} \\
\tilde{f}_{1}(b_0-\delta_n) & \mbox{if $c\in[b_0-\delta_n,b_0]$,}
\end{array}
\right. \nonumber\\
&\hat{f}_{0}(c)=\left\{
\begin{array}{ll}
\tilde{f}_{0}(a_{0}+\delta_n) &\mbox{if $c\in [a_{0},a_{0}+\delta_n]$,} \\
\tilde{f}_{0}(c) & \mbox{if $c\in[a_{0}+\delta_n,b_0-\delta_n]$,} \\
\tilde{f}_{0}(b_0-\delta_n) & \mbox{if $c\in[b_0-\delta_n,b_0]$.}
\end{array}
\right. \label{eq: estimators}
\end{align}

It is well known that the estimator
$\nabla_\beta\widetilde{\TP}(c_{\ell,0},\hat{\beta})$ is in
general inconsistent around the boundary points $a_{0}$ and
$b_{0}$. Therefore, we modify
$\nabla_\beta\widetilde{\TP}(a_{0},\hat{\beta})$ around
the boundary points to obtain uniformly consistent estimators for
$\nabla_\beta \FP(c,\beta^*)$. This method is also used in Donald,
Hsu and Barrett (2012), and Donald and Hsu (2014). Same comment
applies to $\nabla_\beta \widetilde{\FP}(c,\hat{\beta})$, $\tilde{f}_{0}(c)$ and $\tilde{f}_{1}(c)$.
Note that we introduce another $\delta_n$ for $\tilde{f}_{0}(c)$ and $\tilde{f}_{1}(c)$ and this is needed to account for the fact that $\hat{f}_0(c)$ is in the denominator and we need to control its convergence more carefully. We make the
following assumptions on $K(u)$ and $h$.

\begin{assumption}
Assume that $K(u)$ is non-negative and has support $[-1,1]$, $K(u)$
is symmetric around 0 and is continuously differentiable of order 1,
and the bandwidth $h$ satisfies $h\rightarrow 0$, $nh^4\rightarrow
\infty$ and $nh/\log n\rightarrow \infty$ when $n\rightarrow
\infty$.
 \label{assn: kernel}
\end{assumption}

\begin{assumption} 
For any given value of $x$, $G(x,\beta)$ is twice continuously differentiable w.r.t.\ $\beta$ on $B^*(r)$ for some $r>0$ with $\sup_{\beta\in B^*(r),x\in\mathcal{X}}|\partial_{jj}G(x,\beta)|\le M$ almost surely for some $M>0$.
 \label{assn: conditions gradients}
\end{assumption}

\begin{assumption} \label{assn: conditions density 0}
The conditional distribution of $G(X,\beta^*)$ given $Y=0$ has compact support $[a_0,b_0]$ and a twice continuously differentiable probability density function $f_0(c)>0$ satisfying that 
$f_0(c)\geq M\cdot (\min\{c-a_0, b_0-c \})^K$ for some positive integer $K$.
\end{assumption}

\begin{assumption} \label{assn: conditions density 1}
The conditional distribution of $G(X,\beta^*)$ given $Y=1$ has compact support $[a_1,b_1]$ which is a subset of  $[a_0,b_0]$ and a twice continuously differentiable probability density function.  In addition,
$\sup_{c\in[a_0,b_0]}|f_1(c)/f_0(c)|\leq M$ for some $M>0$. 
\end{assumption}

\begin{assumption} \label{assn: delta_n}
Let $\delta_n>0$, $\delta_n\rightarrow 0$ and $\delta_n n^{\iota}\rightarrow \infty$ for any $\iota>0$.  
\end{assumption}

\begin{lemma}\label{lemma: uni consistent estimators}
Suppose that Assumptions \ref{assn: iid}, \ref{assn: beta-est},
\ref{assn: kernel}, \ref{assn: conditions gradients}, \ref{assn: conditions density 0},
 \ref{assn: conditions density 1} and  \ref{assn: delta_n} hold.  
Then the estimators
defined in (\ref{eq: estimators}) satisfy Assumption \ref{assn: consistent estimator}. 
\end{lemma}

\paragraph{Proof of Lemma~\ref{lemma: uni consistent estimators}} 
The proof for uniform consistency of $\nabla_\beta
\widehat{\FP}(c,\hat{\beta})$, $\nabla_\beta
\widehat{\FP}(c,\hat{\beta})$, $\hat{f}_0(c)$ and $\hat{f}_1(c)$ follows the same arguments in 
Donald, Hsu and Barrett (2012), and Donald and Hsu (2014), so we omit it for brevity.  We focus on the uniform consistency of $\hat{f}_1(c)/\hat{f}_0(c)$.
Note that under Assumption \ref{assn: kernel}, we have
\begin{align*}
\sup_{c\in[a_0+\delta_n,b_0-\delta_n]}|\hat{f}_0(c)-{f}_0(c)|=o_p(n^{-1/4}).
\end{align*}
By Assumption \ref{assn: conditions density 0}, we have 
${f}_0(c) \geq M \delta_n^K$ for all $c\in[a_0+\delta_n,b_0-\delta_n]$ and it follows that
uniformly over $[a_0+\delta_n,b_0-\delta_n]$
\begin{align*}
\frac{\hat{f}_0(c)}{f_0(c)}=1+ \frac{\hat{f}_0(c)-{f}_0(c)}{{f}_0(c)}\rightarrow_p 1
\end{align*}
because
\begin{align*}
\Big|\frac{\hat{f}_0(c)-{f}_0(c)}{{f}_0(c)}\Big|\leq \frac{o_p(n^{-1/4})}{ M \delta_n^K}=o_p(1)
\end{align*}
by Assumption \ref{assn: delta_n}.
Next,  uniformly over $[a_0+\delta_n,b_0-\delta_n]$
\begin{align*}
\frac{\hat{f}_1(c)}{\hat{f}_0(c)}
=\frac{\hat{f}_1(c)}{f_0(c)} \cdot \frac{{f}_0(c)}{\hat{f}_0(c)}
=\frac{\hat{f}_1(c)}{f_0(c)} + o_p(1)
=\frac{{f}_1(c)}{f_0(c)} + \frac{\hat{f}_1(c)-{f}_1(c)}{f_0(c)}+ o_p(1)
=\frac{{f}_1(c)}{f_0(c)}+ o_p(1).
\end{align*}
because ${(\hat{f}_1(c)-{f}_1(c))}/{f_0(c)}=o_p(1)$.  Finally, by the fact that ${f}_1(c)/f_0(c)$ is continuous on $[a_0,b_0]$ and $\delta_n\rightarrow 0$, it follows that
\begin{align*}
\sup_{c\in[a_0,b_0]}\left|\frac{\hat{f}_{1}(c)}{\hat{f}_{0}(c)}-\frac{f_{1}(c)}{f_{0}(c)}\right|=o_p(1).
\end{align*} 
Finally, Lemma \ref{lm: uniform quantile} shows that $\sup_{t\in T}|\widehat{c}_t-c_t|=o_p(1)$.

}%END OF FOOTNOTESIZE

\end{document}